# Characterization of large area avalanche photodiodes in X-ray and VUV-light detection

L.M.P. Fernandes[1], F.D. Amaro[1], A. Antognini[2], J.M.R. Cardoso[1], C.A.N. Conde[1], O. Huot[3], P.E. Knowles[3], F. Kottmann[4], J.A.M. Lopes[1], L. Ludhova[3,4], C.M.B. Monteiro[1], F. Mulhauser[3], R. Pohl[2,4], J.M.F. dos Santos[1], L.A. Schaller[3], D. Taqqu[4], J.F.C.A. Veloso[1]

[1] *Physics Department, University of Coimbra, P-3004-516 Coimbra, Portugal*

[2] *Max-Planck Institut für Quantenoptik, D-85748 Garching, Germany*

[3] *Physics Department, University of Fribourg, CH-1700 Fribourg, Switzerland*

[4] *Paul Scherrer Institute, CH-5232 Villigen PSI, Switzerland*

The present manuscript summarizes novel studies on the application of large area avalanche photodiodes (LAAPDs) to the detection of X-rays and vacuum ultraviolet (VUV) light. The operational characteristics of LAAPDs manufactured by Advanced Photonix Inc. were investigated for X-ray detection at room temperature. The optimum energy resolution obtained in four investigated LAAPDs with active areas between 80 and 200 mm$^2$ was found to be in the range 10-18% for 5.9 keV X-rays. The observed variations are associated to the dark current differences between the several LAAPDs. Moreover, the LAAPD simplicity, compact structure, absence of entrance window and high counting rate capability (up to about $10^5$/s) turn it out to be useful in diverse applications, mainly low-energy X-ray detection, where LAAPDs selected for low dark current are able to achieve better performance than proportional counters. LAAPDs have been also investigated as VUV photosensors, where they present advantages compared to photomultiplier tubes. Since X-rays are used as reference in light measurements, the gain non-linearity between X-rays and VUV-light pulses has been investigated. The ratio between 5.9 keV X-rays and VUV light gains decreases with gain. Variations of 10 and 6% were observed for VUV light produced in argon (~128 nm) and xenon (~172 nm) for a gain 200, while for visible light (~635 nm) a variation lower than 1% was measured. The effect of temperature on the LAAPD performance was investigated for X-ray and VUV-light detection. Relative gain variations of about -5% per ºC were observed for the highest gains. The excess noise factor was found to be independent on temperature, presenting values of 1.8 and 2.3 for gains of 50 and 300, respectively. The energy resolution variation with temperature is not related to the excess noise factor, being mainly associated to the dark current. LAAPDs have been tested under intense magnetic fields up to 5 T. Their response in X-ray and visible-light detection practically does not vary with the magnetic field intensity while for 172 nm VUV light a significant amplitude reduction of more than 20% was observed.

**CONTENTS**

## 1. INTRODUCTION

Avalanche photodiodes (APDs) are monolithic devices made of silicon p-n junctions. They have been used as radiation detectors in an increasing number of applications due to their compact structure, simple operation, low power consumption and sensitivity to different radiation types. APDs are able to detect light in the whole visible spectrum, from the infrared to the vacuum ultraviolet (VUV) regions, and X-rays with energy up to about 25 keV.

In the last years, significant advances in the development of large-area avalanche photodiodes (LAAPDs) triggered the characterization of different commercially available APDs [1-4]. They have been used mainly as optical photosensors coupled to scintillators for X-ray, γ-ray and particle detection in applications such as the electromagnetic calorimeter of the Compact Muon Solenoid (CMS) experiment at the Large Hadron Collider (LHC) [2,3], positron emission tomography (PET) instrumentation [5,6] and nuclear physics [1,4].

More recently, windowless LAAPDs with a spectral response extended down to the VUV region (~120 nm) have been developed by API (Advanced Photonix Inc., 1240 Avenida Acaso, Camarillo, CA 93012, USA). As a result of their enhanced quantum efficiency, these LAAPDs can replace photomultiplier tubes (PMTs) or CsI-based photosensors in applications where the high gain is not the most important parameter, such as the detection of primary and secondary scintillation in rare gases in gas proportional scintillation counters [7,8].

The feasibility of using LAAPDs as X-ray detectors has been previously demonstrated [9,10]. Although their use for X-ray detection in the energy range 0.5-20 keV has been suggested [11-13], low-energy X-ray detection techniques with APDs were mainly developed to measure the number of charge carriers in light measurements, using X-rays as a reference [11,13-15], resulting in a straightforward process to evaluate the number of photons interacting in the photodiode.

The objective of our studies about LAAPDs was to investigate their response as direct X-ray detectors for X-ray spectrometry applications as well as VUV photosensors in gas proportional scintillation counters (GPSCs). Studies with X-rays include the operational characteristics of LAAPDs from API at room temperature [16], the application of pulse rise-time discrimination techniques to LAAPD signals [17], the behaviour under intense magnetic fields (up to 5 Tesla) [18,19] and the effect of temperature on the LAAPD gain and performance, mainly the energy resolution contributions [20-22].

Concerning the use of LAAPDs as VUV photosensors in GPSCs [7,8,23], their response to VUV-light pulses resulting from scintillation of argon (~128 nm) and xenon (~172 nm) was investigated. The LAAPD operational characteristics for VUV-light detection were studied, including the minimum detection limit and statistical fluctuations in VUV photon detection, the gain non-linearity between X-rays and VUV light [24], the gain dependence on temperature [25] and the behaviour under intense magnetic fields [18].

## 2. THE AVALANCHE PHOTODIODE

For a long time, only photomultiplier tubes (PMTs) and conventional photodiodes provided quantitative detection in the whole visible light spectrum. The avalanche photodiode combines the benefits of both photosensors since it is a silicon photodiode with internal gain. The gain is however significantly lower than the one obtained with a PMT, reaching less than $10^3$. The gain is obtained by applying a high reverse bias voltage to the photodiode which creates an intense electric field inside the APD.

Non-uniformity has been a major drawback in the manufacture of LAAPDs limiting their applications. However, LAAPDs from API have been developed with improved spatial uniformity, delivering higher gains at lower bias voltages [14]. Comparing to others, LAAPDs from API present higher quantum efficiency and lower noise levels [26]. The basis of the API technology is the development of silicon crystals with n-type doping obtained by neutron transmutation, with more uniform resistivity, which provides larger avalanche regions with more moderate electric fields and therefore lower dark current. According to API, only crystals with resistivity variations lower than 5% are used in the manufacture of LAAPDs.

When a reverse bias voltage is applied to the photodiode, the maximum electric field is reached around the p-n junction and at higher voltages the field could be so strong causing rupture of the junction edges. To prevent this problem, LAAPDs from API have bevelled edge geometry (Fig. 1), which reduces the electric field strength in the junction edges.

### 2.1. Structure and operation

Figure 1 shows the structure of a bevelled edge LAAPD, showing the electric field profile inside its volume. When a high voltage is applied to the LAAPD, only a small region of the p-layer, near the photodiode surface, remains undepleted - the drift region (i). This region shows a residual electric field of about 50 V/cm [27] due to the high acceptor concentration. In the depleted p-region (ii), the electric field increases with depth until reaching a maximum, about $1.8 \times 10^5$ V/cm [27], near the p-n junction and decreases in the depleted n-region. Light photons, X-rays or charged particles absorbed in the p-region are converted into electron-hole pairs. The resulting primary electrons are driven to the p-n junction by the electric field. Around the junction, they obtain a sufficient amount of energy to produce new electron-hole pairs by impact ionization, originating an avalanche process in the multiplication region (iii). Charge gains of a few hundred are typical and increase exponentially with the applied voltage, resulting in a significant improvement of the signal to-noise ratio. Detailed operation principles of the APD have been presented in the literature [9,14,28].

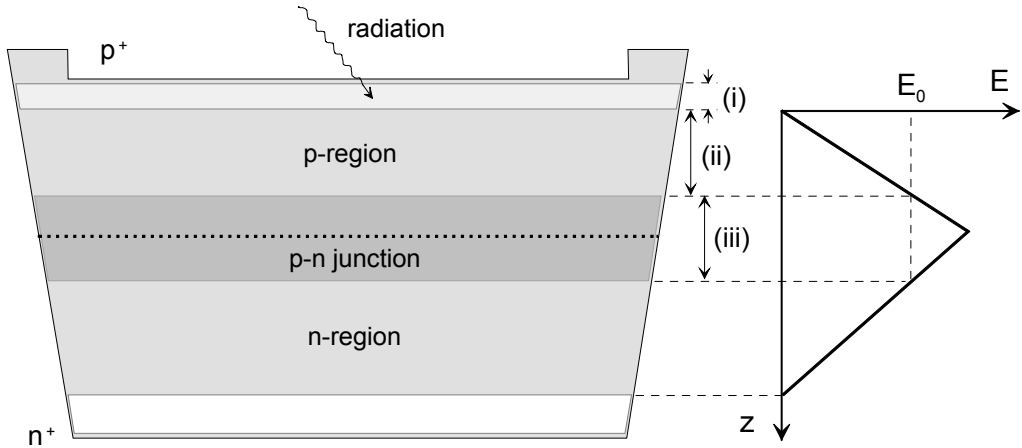

Fig. 1. Schematic of a bevelled edge avalanche photodiode and the electric field (E) profile inside its volume. The active region of the APD can be divided in three different parts: the drift region (i), the depleted p-region (ii) and the multiplication region (iii), where the electric field is higher ($E > E_0$).

The silicon active thickness extends from the APD surface to the multiplication region edge in the n-region side. As shown in Fig. 1, this active region can be divided in three different zones with different response to X-rays: the drift region (i), the depleted p-region where the electric field is not enough to provide electron multiplication (ii) and the multiplication region (iii). For LAAPDs from API, the active region is about 12 μm thick, while the multiplication region is about 7 μm thick for intermediate bias voltages (1600 to 1800 V).

Due to the weak electric field in region (i), X-rays there absorbed produce electrons that go slowly towards the depletion region edge and, consequently, the resulting pulses have longer time responses. In this region, electrons can be captured for long periods of time, from tens to thousands of ns [28], and the output pulses may present lower amplitude, depending on the integration time. X-rays absorbed in region (ii) generate fully amplified pulses with faster time response compared to X-rays absorbed in the drift region. The primary electrons are rapidly carried to the high electric field region, undergoing an average amplification equal to the APD gain. X-rays absorbed in the multiplication region produce electrons that will be partially amplified, originating lower amplitude pulses with faster time responses.

X-rays with the same energy can be absorbed in any of the regions (i), (ii) or (iii). The amplitude distribution of the resulting pulses deviates from a Gaussian curve due to a low-energy tail associated to lower amplitude pulses generated in regions (i) and (iii). Since the time response is also different in these regions, pulse rise-time discrimination may improve the APD performance.

When a voltage is applied to a photodiode in order to polarize the p-n junction, a low-intensity current, typically a fraction of μA, is observed. This dark current has its origin in the detector volume and surface. The volumetric dark current results from the continuous generation of charges (minority carriers) in both sides of the junction, which are conducted through it, and the thermal generation of electron-hole pairs in the depletion region, which increases with the volume and decreases by cooling. The superficial dark current is generated in the p-n junction edges due to high voltage gradients nearby. Since dark current is a noise source and increases considerably with temperature, the electronic noise level can be reduced by cooling the APD, improving the energy resolution in X-ray detection [20,25].

In VUV-light detection, the photon absorption takes place in the first silicon layers due to the reduced absorption length of those photons, about 5 nm [29]. For visible light, the penetration depth is about 1 μm and the photon absorption takes place deeper compared to VUV-light but still before the multiplication region. In these cases, the LAAPD response does not vary significantly from event to event. In opposite, X-rays can interact in different regions of the LAAPD, generating pulses with different amplitude and time responses.

In X-ray detection, the LAAPD efficiency depends on the X-ray absorption length in silicon, which varies with energy according to Fig. 2. The majority of absorbed X-rays originate fast electrons which total energy is approximately equal to the incident photon energy. This electron is stopped in successive collisions and originates in average an electron-hole pair per each 3.62 eV of energy deposited in silicon [30]. Each X-ray absorbed in the APD can be considered as a local deposition of E/ε electron-hole pairs, being E the incident X-ray energy and ε the average energy required to create an electron-hole pair (3.62 eV).

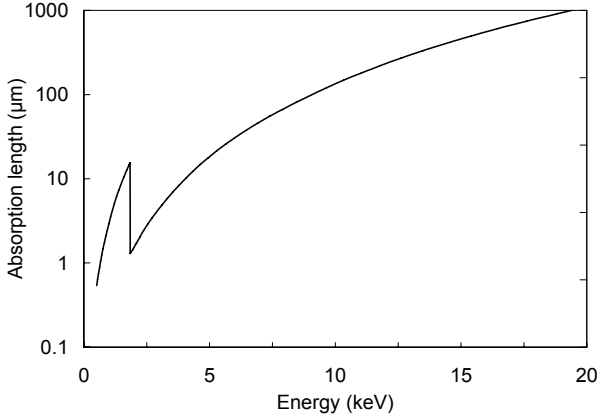

Fig. 2. Average absorption length in silicon as a function of the X-ray energy [31].

The penetration depth in silicon is about 30 μm for 6 keV X-rays and increases rapidly with the energy, reaching about 70, 130 and 450 μm for X-ray energies of 8, 10 and 15 keV, respectively. This dependence affects the APD detection efficiency, defined as the ratio between the number of fully amplified pulses obtained at the APD output and the number of photons incident in the APD surface. The efficiency decreases rapidly for energies above 6 keV, being about 45, 25, 8 and 4% for 8, 10, 15 and 20 keV energies, respectively [11, 28].

*2.2. Intrinsic energy resolution*

The energy resolution associated to radiation detection in avalanche photodiodes is determined by several factors:

- statistical fluctuations associated to the number of electron-hole pairs created in silicon and the avalanche process ($\Delta E_S$);
- gain non-uniformity in the APD detection volume ($\Delta E_U$);
- detector noise (resulting from dark current) and electronic system noise ($\Delta E_N$).

The total broadening ($\Delta E$) in the energy distributions of APD pulses is the quadratic addition of the three contributions:

$$\Delta E^2 = \Delta E_S^2 + \Delta E_U^2 + \Delta E_N^2 \quad (1)$$

The output signal variance associated to the statistical contribution, in number of primary electrons, is given by [30,32]:

$$\sigma_S^2 = \sigma_n^2 + N(F-1) \quad (2)$$

being $N$ the number of primary electrons, $\sigma_n^2$ its variance and $F$ the excess noise factor. $F$ is related to the variance of the electron avalanche gain, $\sigma_A^2$, according to:

$$F = 1 + \sigma_A^2 / G^2 \quad (3)$$

Due to the discrete nature of the multiplication process, as a result of electron avalanche fluctuations F is higher than 1 and varies with the gain (G).

There is a clear difference between light and X-ray detection. In particular, the non-uniformity contribution is negligible in light detection if the whole APD area is irradiated since the final pulse results from the average response to the entire number of photons interacting in the silicon. In X-ray detection, each pulse is locally created and the final distribution is affected by the position gain variation.

For light pulse detection, the variance of the number of primary electrons is described by Poisson statistics, $\sigma_n^2 = N$. The statistical error, in number of primary electrons, then is:

$$\sigma_S^2 = NF \quad (4)$$

If the whole photodiode area is irradiated, the contribution of gain non-uniformity to the peak broadening can be neglected. The intrinsic resolution (the total energy resolution without the noise contribution) in light detection is basically determined by the statistical contribution:

$$R_{int} = 2.36 \frac{\sigma_S}{N} = 2.36 \sqrt{\frac{F}{N}} \quad (5)$$

For X-ray detection, the peak broadening process is more complex. The statistical fluctuations associated to primary electrons are attenuated by the Fano factor $f$. For silicon, $f$ is about 0.12 for 5.9 keV X-rays [33]. The variance of the number of primary electrons is now $\sigma_n^2 = Nf$ and the statistical contribution to the energy resolution is:

$$\sigma_S^2 = N(F + f - 1) \quad (6)$$

For X-rays, the energy resolution can be seriously degraded due to the gain non-uniformity in the detection volume. If $\sigma_U / G$ is the relative standard deviation associated to gain non-uniformity, the intrinsic resolution for X-rays can be described by:

$$R_{int} = 2.36 \sqrt{\frac{F + f - 1}{N} + \left(\frac{\sigma_U}{G}\right)^2} \quad (7)$$

*2.3. Noise contribution and total energy resolution*

The noise contribution to the energy resolution results from two different sources: the detector dark current and the electronic system. Dark current presents two different components. One of them ($I_{DS}$) does not depend on gain and corresponds to the superficial current and to a small fraction of the volumetric current resulting from thermal generation of electron-hole pairs in the n-region, thus non-amplified. The other component ($I_{DV}$) is amplified by the gain and corresponds to the volumetric current resulting from the generation of electron-hole pairs in the p-region. The total current at the APD output is then:

$$I = I_{DS} + G I_{DV} + G I_0 \qquad (8)$$

where $G$ is the APD gain and $I_0$ the non-amplified signal current, corresponding to electron-hole pairs created by the absorbed radiation.

The noise associated to the electronic system is mainly originated in the FET (field effect transistor) at the preamplifier input. Fluctuations in the FET channel current are similar to thermal noise and can be represented by a noise equivalent resistance ($R_{eq}$) in series with the preamplifier input [34].

A detailed noise analysis in avalanche photodiodes has been already presented in the literature [12,35]. If the preamplifier is connected to a linear amplifier with equal differentiation and integration constants $\tau$, the electronic noise contribution to peak broadening (in units of energy) is:

$$\Delta E_N^2 = \left(2.36 \frac{e\varepsilon}{qG}\right)^2 \left[\frac{k_B T R_{eq}}{2\tau} C_T^2 + \frac{\tau q}{4}(I_{DS} + I_{DV} G^2 F)\right] \qquad (9)$$

being q the electron charge, $e \cong 2.718$ the number of Nepper, $k_B$ the Boltzmann constant ($1.38 \times 10^{-23}$ J/K) and $T$ the temperature (in Kelvin); $C_T$ is the total capacitance at the preamplifier input (including detector and FET input capacitances).

The first term in Eq. (9) describes the electronic system noise associated to the detector and the second term corresponds to the dark current contribution. Both terms depend on the shaping time constants used in the linear amplifier. The noise contribution also depends on the gain and the excess noise factor.

The relationship between $F$ and $G$ has been derived from the McIntyre model considering that photoelectrons are injected next to the p-zone surface [36]:

$$F \cong G k_{ef} + \left(2 - \frac{1}{G}\right)(1 - k_{ef}) \qquad (10)$$

being $k_{ef}$ the effective ratio between the ionization coefficients for holes and electrons. For lower gains, $k_{ef} \ll 1$ and then $F \cong 2 - 1/G$.

For the useful gain range ($G > 30$), the variation of kef with the voltage is very low and considering kef constant is a good approximation. As a result, the dependence of $F$ on $G$ should be approximately linear.

Equation (10) makes it possible to estimate the main operational parameters of the LAAPD, such as the energy resolution and the optimum gain. The total energy resolution is given by the equation:

$$R^2 = R_{int}^2 + \left(2.36 \frac{e\varepsilon}{qEG}\right)^2 \frac{k_B T R_{eq}}{2\tau} C_T^2 + \\ + \left(2.36 \frac{e\varepsilon}{qEG}\right)^2 \frac{\tau q}{4}(I_{DS} + I_{DV} G^2 F) \qquad (11)$$

where $R_{int}$ is the intrinsic resolution, given by Eqs. (5) and (7) for light and X-ray detection, respectively, and $E$ is the energy deposited in silicon by the incident radiation.

The optimum gain is the minimum of the energy resolution, being a solution of equation $\partial R^2 / \partial G = 0$. Since $\partial R / \partial G \cong k_{ef}$ for the APD useful gain region, the optimum gain estimate, $G_{op}$, is given by:

$$G_{op}^3 = \frac{2}{k_{ef}}\left(I_{DS} + 2\frac{k_B T}{q} \frac{R_{eq} C_T^2}{\tau^2}\right) \bigg/ \left(I_{DV} + \frac{4q}{e^2\tau}\frac{E}{\varepsilon}\right) \qquad (12)$$

The optimum gain depends on factors inherent to the APD, as $k_{ef}$ and the dark current, as well as on the preamplifier characteristics and shaping time constants. According to Eq. (12), the optimum gain decreases slightly with the energy deposited by the incident radiation.

The previous parameters can be estimated considering typical values in Eqs. (11) and (12). For X-rays, the intrinsic resolution is given by Eq. (7). The dependence of $F$ on $G$ can be found using Eq. (9). Typical values of the dark current in API avalanche photodiodes, with 16 mm diameter, are $I_{DS} = 100$ nA and $I_{DV} = 0.3$ nA at room temperature (294 K), while $k_{ef} \approx 0.0015$ [37]. The capacitance of the photodiodes in the useful gain range is about 130 pF [27]. Considering the use of a charge sensitive preamplifier with a JFET mutual conductance of 20 mS and an input capacitance of 1 pF, $R_{eq} \cong 33\,\Omega$ [34] and $C_T \cong 130$ pF.

Figure 3 shows the total energy resolution estimated for different X-ray energies as a function of gain, assuming shaping time constants of 200 ns and a non-uniformity of 2% (standard deviation). The figure shows that the optimum gain, corresponding to minimum energy resolution, decreases slightly as the X-ray energy increases, being about 125, 100 and 75 for 2, 6 and 20 keV, respectively, what is obviously in accordance with Eq. (12). For each individual LAAPD, a comparison between theoretical estimates and experimental values can be made and the different contributions to the energy resolution can be evaluated.

The non-uniformity depends on each individual photodiode and the same happens with dark current, which may vary considerably from APD to APD, leading to significant differences on the intrinsic resolution and noise for different photodiodes.

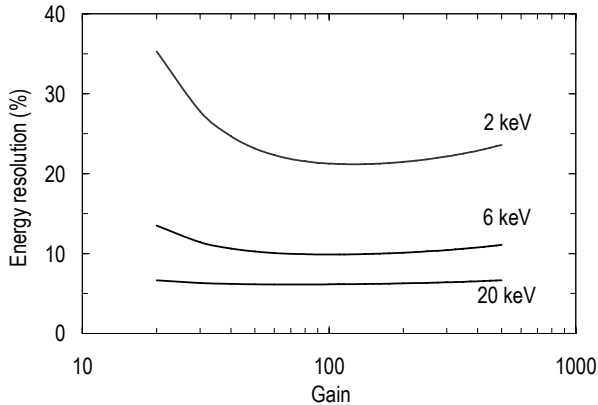

Fig. 3. Estimate of the total energy resolution as a function of gain for X-ray energies of 2, 6 and 20 keV, considering 2% non-uniformity and 200 ns shaping time constants.

## 3. X-RAY DETECTION

The performance characteristics of LAAPDs in X-ray detection were determined first at room temperature, including gain and the dark current effect on the energy resolution. The spatial uniformity was investigated, as well as the counting rate capability and space charge effects. The dependence of the gain and dark current on temperature was also studied and its influence on the LAAPD performance investigated. Finally, the behaviour of LAAPDs under intense magnetic fields (up to 5 T) was also investigated.

### 3.1. Performance characteristics at room temperature

Several windowless UV-enhanced LAAPDs from API were investigated. The characteristics of each LAAPD were evaluated with 5.9 keV X-rays from a $^{55}$Fe source (Mn K$_\alpha$-line). The Mn K$_\beta$-line (6.4 keV X-rays) was efficiently reduced by absorption in a chromium filter. The LAAPDs were shielded from the ambient light. During measurements, room temperature was stabilized at 20 ± 0.3ºC. Table 1 shows the main specifications for each APD according to data sheets.

Table 1. LAAPD specifications just before breakdown.

| LAAPD No. | Diameter (mm) | Voltage (V) | Gain | Dark Current (nA) |
|---|---|---|---|---|
| 1 | 16 | 1840 | 317 | 201 |
| 2 | 16 | 1851 | 312 | 336 |
| 3 | 16 | 1849 | 303 | 499 |
| 4 | 10 | 1873 | 314 | 222 |

X-ray signals produced at the LAAPD output were fed through a low-noise charge preamplifier (Canberra 2004), with a sensitivity of 45mV/MeV, to a linear amplifier (HP 5582A) and stored in a 1024-channel analyzer (Nucleus PCA II).

In order to determine the optimum shaping time constants in the linear amplifier, the LAAPD performance, the energy resolution and minimum detectable energy (MDE) were determined for different shaping constants. Table 2 shows the values obtained with LAAPD 1, polarized with 1830 V, for counting rates of about $10^3$/s. The MDE is determined by the low-energy noise tail in the pulse-height distribution.

Table 2. Energy resolution and minimum detectable energy (MDE) for 5.9 keV X-rays detected in the LAAPD 1 for different shaping time constants in the linear amplifier.

| Time Constant (ns) | 50 | 100 | 200 | 500 | 1000 |
|---|---|---|---|---|---|
| Energy Resolution (%) | 12.6 | 12.7 | 13.2 | 14.6 | 16.7 |
| MDE (keV) | 0.90 | 0.89 | 0.96 | 1.18 | 1.52 |

Table 2 shows that the energy resolution and MDE increase significantly for shaping constants longer than 200 ns. Since the energy resolution measurements have errors of ± 0.1%, optimum shaping constants are in the interval 50-200 ns. Since shorter shaping constants (less than 100 ns) may originate partial signal loss, the choice of shaping time constants between 100 and 200 ns is reasonable.

Using 200 ns shaping constants in the linear amplifier, the LAAPD performance characteristics were determined from a collimated X-ray beam (1 mm diameter), as a function of bias voltage, for the several LAAPDs. The gain was determined by normalizing the signal amplitude to the manufacturer specification at 1400 V (gain 3.5). The energy resolution and MDE are presented in Fig. 4 as a function of gain for LAAPD 4. The best energy resolution is achieved for gains around 50 and degrades at higher gains. The same trend is observed for the MDE but its increase at high gains is less significant.

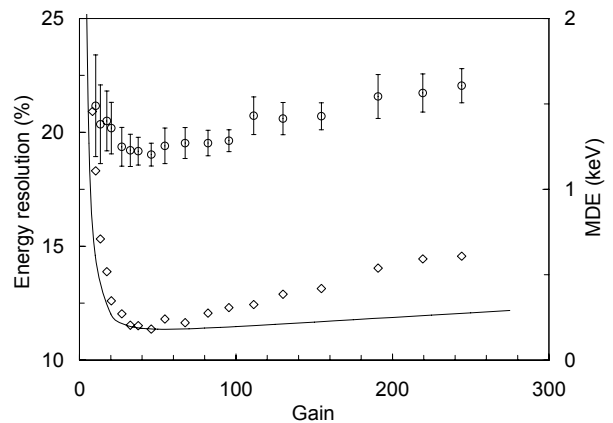

Fig. 4. Energy resolution for 5.9 keV X-rays and minimum detectable energy (MDE) as a function of gain for LAAPD 4. The experimental errors for the energy resolution fall within the symbol size. The curve is an estimate of the energy resolution, Eq. (11).

An estimate of the energy resolution is also presented in Fig. 4. This estimate was determined by Eq. (11) using typical parameters of the LAAPD for $\tau = 200$ ns and $T = 20\,^{\circ}\text{C}$. The intrinsic resolution was assessed by Eq. (7) for 5.9 keV X-rays ($N = 1630$), being $F$ derived from Eq. (10), with $k_{ef} \approx 0.0015$, and assuming a gain non-uniformity $\sigma_U / G = 0.02$ [37].

To determine the noise contribution, the dark current values provided by the manufacturer ($I_D$) were plotted against gain ($G$). Dark current components of $I_{DS} = 24.2$ nA and $I_{DV} = 0.64$ nA were obtained, $I_D = I_{DS} + G I_{DV}$. Taking into account the typical capacitance of the API avalanche photodiodes (65 or 130 pF for LAAPDs with 10 or 16 mm diameter [27]) and the 1 pF FET input capacitance of the preamplifier, the total capacitance in the preamplifier input is determined by the LAAPD ($C_T \cong 65$ pF for LAAPD 4). Since the mutual conductance $g_m$ of the JFET (2N5434 N-Channel JFET Switch, Calogic Corporation) is 20 mS, the noise equivalent resistance in the preamplifier input is $R_{eq} = 2/(3 g_m) \cong 33\,\Omega$.

The difference between experimental and estimated values of the energy resolution is small for gains around the optimum but increase for higher gains. In this region, the resolution is mainly affected by the dark current, which was not experimentally measured due to a limitation of the experimental system (below 1 µA). Since the dependence of the dark current on gain deviates from a linear relationship for high gains, this may result in larger differences between the energy resolution measurements and the estimate.

According to Eq. (12), the optimum gain estimate is about 80 for LAAPD 4 and does not depend on the assumed non-uniformity. The measured optimum gain is around 50. However, the measured energy resolution varies between 11.4 and 12.0% for gains between 30 and 80. In this case, we may conclude that the choice of the optimum gain is not critical since the LAAPD performance does not vary significantly in a reasonable range of gains.

Table 3. Optimum performance characteristics of several LAAPDs for 5.9 keV X-rays.

| LAAPD No. | Optimum Gain | M.D.E. (keV) | E. Resolution (1mm diameter) | E. Resolution (full area) |
|---|---|---|---|---|
| 1 | 72 | 0.9 | 10.3% | 12.3% |
| 2 | 53 | 1.1 | 11.8% | 14.9% |
| 3 | 52 | 2.2 | 17.9% | 18.8% |
| 4 | 46 | 1.2 | 11.4% | 12.3% |

Table 3 presents the optimum operational characteristics (gain, minimum detectable energy and energy resolution) of each tested LAAPD for 5.9 keV X-rays. The energy resolution was determined for 1 mm collimation of the X-ray beam and for full area illumination. In the first case, the best energy resolution was found to vary from 10 to 18% for the different LAAPDs. For full area illumination, the energy resolution is worst due to the gain non-uniformity in the silicon.

Figure 5 shows typical pulse-height distributions for 5.9 keV X-rays detected by the different LAAPDs. The X-ray peak departs from the gaussian shape, presenting a tail towards the low energy region. This tail results from X-ray interactions in the multiplication or drift regions of the LAAPD (Fig. 1), as described before. The figure shows clearly the correspondence between the photodiode dark current (Table 1) and the MDE and peak broadening. Higher dark current values result in reduced performance for both energy resolution and MDE.

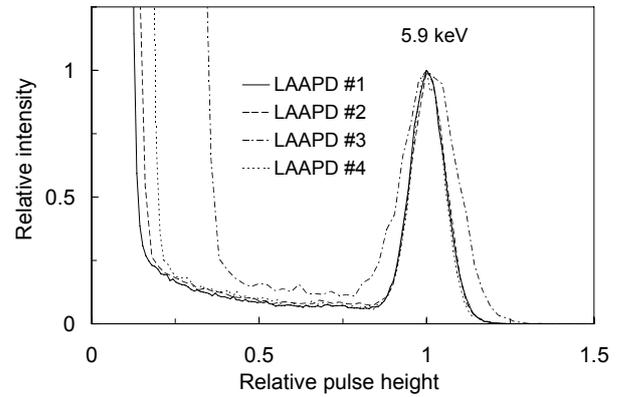

Fig. 5. Typical pulse-height distributions for 5.9 keV X-rays detected by the LAAPDs described in Table 1, for the optimum gain of each LAAPD.

### 3.2. Spatial uniformity

Non-uniform silicon resistivity of the LAAPD results in gain fluctuations due to local electric field variations. As a result, the LAAPD response to incident X-rays depends on their interaction position in the photodiode surface. The spatial uniformity was studied for two different LAAPDs using 5.9 keV X-ray collimated beams (1 mm diameter) positioned at 17 equally distributed points over the photodiode area [16]. The distribution of pulse amplitudes obtained in the different points, normalized to the average value, was determined for the same bias voltage, corresponding to a gain around the optimum. For all LAAPDs, the relative standard deviation due to non-uniformity was found to be in the 2-3% range. No correlation between dark current and non-uniformity was found.

### 3.3. Counting rate capability

The LAAPD pulse amplitude and energy resolution was investigated as a function of the counting rate for 5.9 keV X-ray beams (8 mm diameter). Figure 6 shows the results obtained for LAAPD 4 using 100 ns shaping time constants. The LAAPD is able to work at counting rates as high as $10^4$/s without significant degradation of the relative amplitude and energy resolution. For $4\times10^4$ counts/s, a 3% pulse amplitude reduction and an energy resolution degradation from 12% to 13% are observed. The same trend was observed for the other LAAPDs.

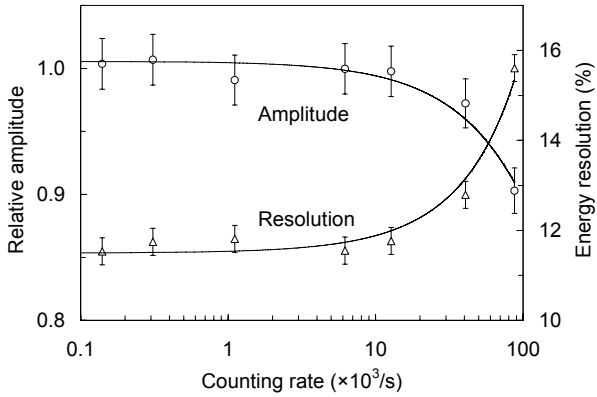

Fig. 6. Relative amplitude and energy resolution for 5.9 keV X-rays as a function of the counting rate in LAAPD 4, for shaping time constants of 100 ns. The lines are exponential fits to the amplitude and energy resolution data.

For higher counting rates, a distortion of the peak shape is observed due to pulse pile-up. Shorter time constants provide high counting rate capability without significant degradation of the energy resolution and minimum detection limit (Table 2).

### 3.4. Space charge effects

Space charge effects may take place when high signal current densities are produced by absorbed X-rays at high gains, resulting in an electric field reduction and local heating in the avalanche region. As a result, the LAAPD response is non-linear. The detector linearity was investigated by comparing the pulse amplitudes for 5.9 and 22.1 keV X-rays, emitted by $^{55}$Fe and $^{109}$Cd radioactive sources, as a function of gain. Figure 7 shows the ratio between the normalized pulse amplitudes for 22.1 and 5.9 keV X-rays absorbed in LAAPD 4. The ratio decreases with increasing gain. However, for gains up to 100 this effect is negligible since less than 1% variation is found. For gains of about 200 and 250, the decrease is about 3 and 6%, respectively.

The LAAPD non-linearity was also investigated for higher energy X-rays using a $^{241}$Am source, which emits 59.6 keV X-rays as well as 13.9 and 17.6 keV X-rays (from the Np fluorescence lines $L_\alpha$ and $L_\beta$). The amplitude ratio between 59.6 and 17.6 keV X-ray pulses was monitored as a function of gain and variations of 6, 10 and 17% were observed for gains of 50, 100 and 200, respectively. The gain non-linearity between X-rays with different energy is more significant for higher energies since larger avalanches are created, originating higher current densities in the LAAPD.

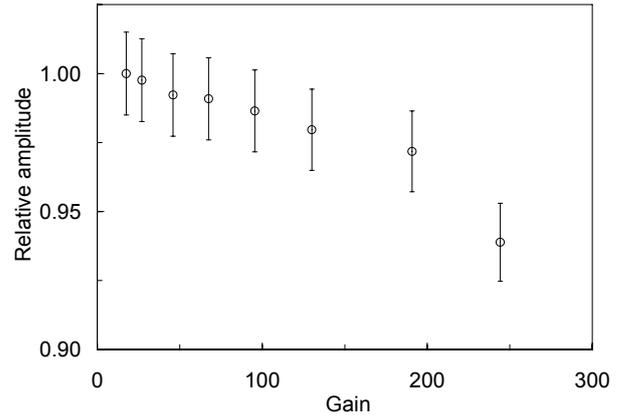

Fig. 7. Relative amplitude of pulses produced in LAAPD 4 by 22.1 and 5.9 keV X-rays as a function of the LAAPD gain.

### 3.5. Effect of temperature on the LAAPD gain and performance

The strong dependence of the LAAPD gain on temperature may be a drawback for many applications, requiring temperature control and stabilization during measurements. In many applications, this is not possible and temperature corrections can be made from the knowledge of the gain variation with temperature. On the other hand, the LAAPD dark current is strongly reduced as the temperature decreases and the operation of LAAPDs at reduced temperatures may result in improved performance [11,13,20,22,38].

The performance of standard LAAPDs from API was investigated for X-ray detection as a function of temperature. The LAAPDs were operated in a light-tight box to shield them from the ambient light. Two different cooling systems were used: a Peltier module providing minimum operation temperatures of -5ºC and a continuous flow of cooled gaseous nitrogen providing LAAPD cooling down to -40ºC. The temperature was stabilized within ± 0.1ºC and ± 0.5ºC, respectively. The LAAPDs were irradiated with an X-ray beam from a 55Fe source. In the second case, light pulses from a LED were also used for comparison purposes. The LED, with emission peak at 635 nm, was connected to a 50 Ω resistance and supplied by a LED-pulser giving pulses with amplitude up to 10 V and width from 6 to 500 ns. A light guide was used to carry the light pulses to the LAAPD window.

The unitary gain was assessed with 100 ns long visible-light pulses for voltages lower than 500 V [22]. The LAAPD gain was then determined for 5.9 keV X-rays by normalization to visible-light measurements. Figure 8 shows the gain for X-rays as a function of bias voltage for different temperatures. For the same voltage, the gain increases as temperature decreases.

Figure 9 shows the gain for 5.9 keV X-rays as function of temperature for different bias voltages, obtained with the Peltier cooled LAAPD (a) and with the LAAPD cooled by nitrogen (b). For each voltage, the relative variation of the gain is almost constant for temperatures above -15ºC.

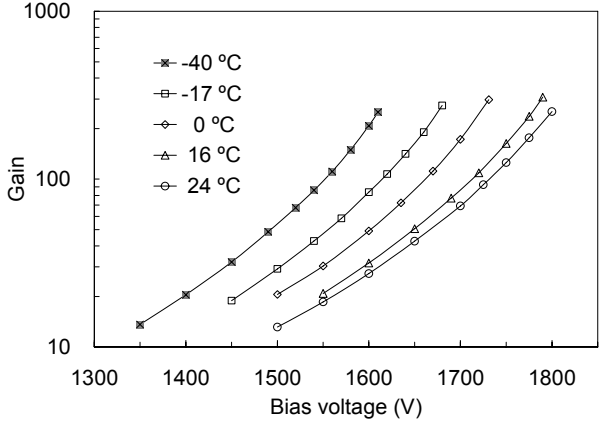

Fig. 8. LAAPD gain for 5.9 keV X-rays as function of the reverse bias voltage, for different temperatures from -40 up to 24ºC. The lines are not fits to the data; they just pass through the points.

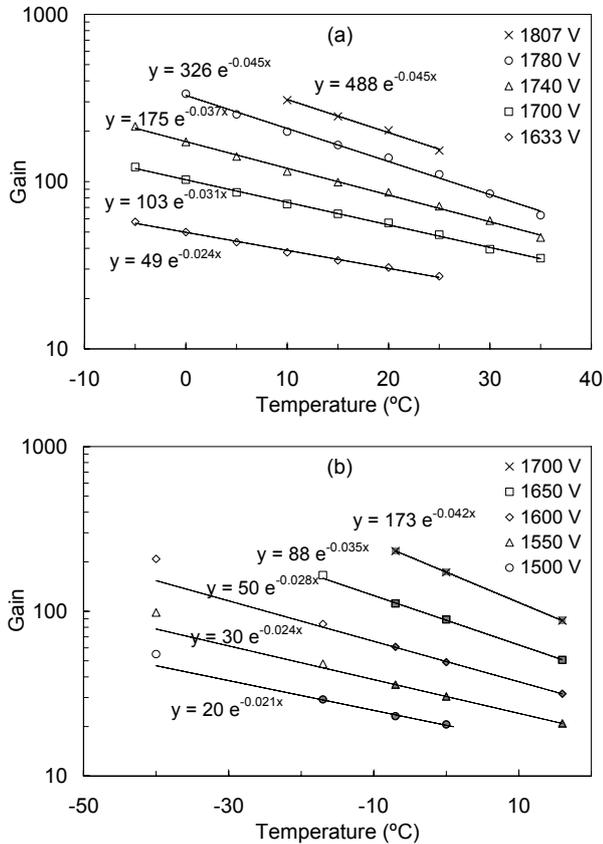

Fig. 9. Gain for 5.9 keV X-rays obtained in two different LAAPDs, cooled by a Peltier module (a) and by cold gaseous nitrogen (b), as a function of temperature for different bias voltages. The lines are exponential fits to the data.

The variation with temperature of the LAAPD gain for X-rays was determined for the first time; only results for visible light have been reported in the literature. For both LAAPDs investigated, the relative gain variation increases with voltage, reaching about -4.5% per ºC for gains around 300. These values are significantly larger than the one reported by the manufacturer for visible light, about -3% per ºC for the maximum bias voltages [27].

The gain non-linearity between X-ray and visible-light signals as a function of temperature was also investigated. The LAAPD was simultaneously irradiated with X-rays from a $^{54}$Mn source and visible-light pulses from the LED. The LED intensity was chosen to produce pulses with amplitude equivalent to the one produced by 9 keV X-rays. Figure 10 shows the gain ratio between 5.4 keV X-rays and visible-light pulses as a function of light gain. The variation is smaller than 2% for gains up to 300. Moreover, the gain non-linearity is even more reduced as the temperature decreases, being smaller than 0.4% and 0.2% for 0 and -12ºC, respectively, for gains up to 200. This dependence on temperature may be related to the penetration depth. Since for visible-light the average absorption length in silicon is about 1 μm [29], practically all photons leave their energy before the APD multiplication region. For 5.4 keV X-rays, the average absorption length is 22 μm (Fig. 2), approximately equal to the distance between the LAAPD surface and the multiplication region. Thus, a number of X-rays are absorbed in that region and partial signal amplification takes place. As temperature increases, the avalanche zone also increases and more pulses with lower amplitude are produced. As a result, the ratio between X-ray and visible-light gains decreases.

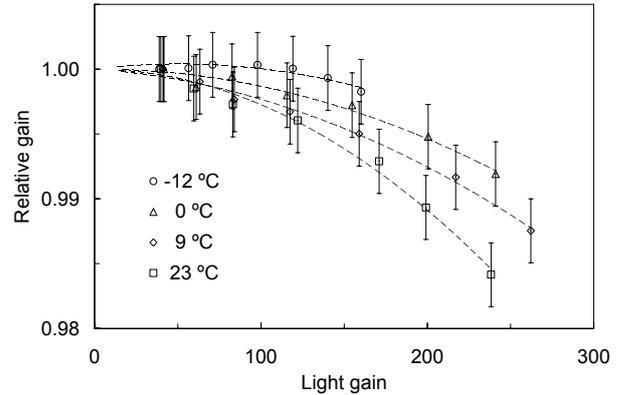

Fig. 10. Relative gain for 5.4 keV X-rays and visible-light pulses as function of the light gain, for different temperatures. The doted lines are used to guide the eyes.

The LAAPD non-linear response for X-rays with different energies was also investigated for different operation temperatures by irradiating the LAAPD with a $^{57}$Co source, which emits 14.4 and 6.4 keV X-rays [22]. The results are consistent with those obtained for room temperature; the pulse amplitude ratio between 14.4 and 6.4 keV X-rays decreases with gain, reaching a 3% variation at gains of about 300. In this case, no dependence on temperature was found. This happens because for X-rays with energy above 6 keV the average absorption length is larger than the distance between the LAAPD surface and the multiplication region. Thus, the effect of temperature on the avalanche size does not vary significantly with energy. The deviation from linearity results only from space charge effects, which obviously increases with energy.

The LAAPD performance was also investigated for different temperatures. Figure 11 shows the MDE (a) and the energy resolution for 5.9 keV X-rays (b) as a function of gain for different temperatures. The measurements were made for full illumination of the LAAPD active area with X-rays from a $^{55}$Fe source at counting rates of about $10^2$/s. Figure 11 shows that the LAAPD performance tends to improve with decreasing temperature. However, below 0ºC the improvement is not significant.

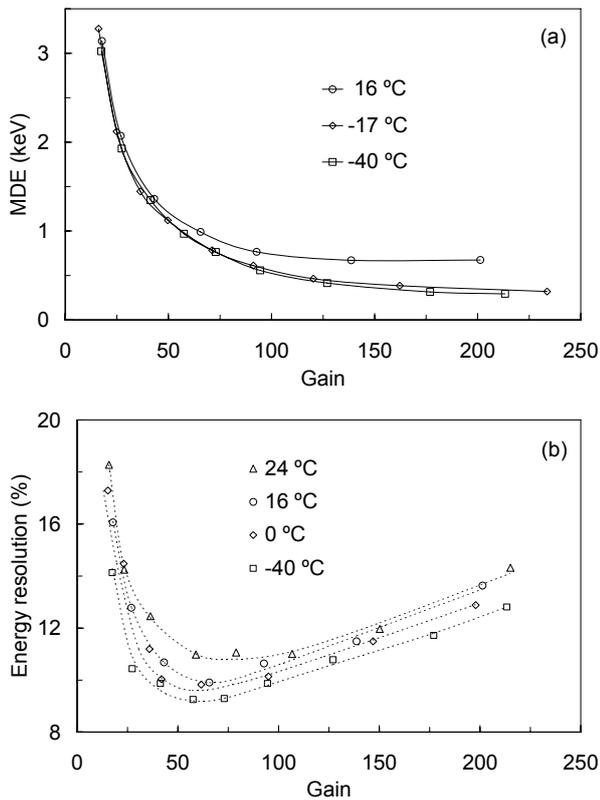

Fig. 11. Minimum detectable energy (a) and energy resolution for 5.9 keV X-rays (b) as a function of gain, for different temperatures. In (b), the doted lines are used to guide the eyes.

The MDE initially presents a fast decrease with gain, stabilizing for high gains. The LAAPD is useful for X-ray detection down to about 0.7 keV at room temperature or even down to 0.3 keV when it is cooled down to below 0ºC. These values correspond to a minimum number of primary electron/hole pairs of about 200 and 80, respectively, to produce a signal above the noise level.

As seen in Fig. 11 (b), the optimum energy resolution is obtained for gains in the 60-80 range. The optimum gain does not depend significantly on temperature. The energy resolution is better than that observed in proportional counters and improves from about 11% to 9.6 and 9.2% as the temperature decreases from 24ºC to 0 and -40ºC. Below 0ºC, the improvement is not significant. An energy resolution as good as 11% for full area illumination was not obtained before at room temperature but this is a different LAAPD, with better dark current characteristics.

To understand the energy resolution variation with temperature, some contributions were determined as a function of gain, namely the dark current and the excess noise factor. Both these parameters contribute to the electronic noise, while the excess noise factor itself also contributes to the intrinsic resolution, according to Eqs. (7) and (9). In addition do these contributions, the significant degradation of the energy resolution for high gains is also related to an incorrect fit of two different Gaussian curves to the Mn $K_\alpha$ and $K_\beta$ lines, which are superimposed. In fact, when only 5.9 keV X-rays from the Mn $K_\alpha$-line were detected, the energy resolution degradation observed at high gains was much smaller [20].

Figure 12 shows the dark current as a function of gain, measured for different temperatures. As expected, the dark current variation with gain is approximately linear. For a given gain, the dark current is reduced by more than one order of magnitude as the operation temperature is reduced from 24ºC to 0ºC, which is in accordance with the manufacturer specification [27].

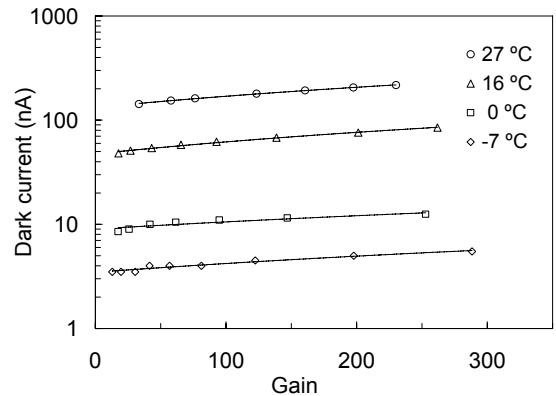

Fig. 12. LAAPD dark current as a function of gain for different temperatures. The lines are linear fits to the points.

The excess noise factor was determined by the simultaneous detection of X-rays and visible-light pulses [21]. In the method used, the excess noise factor is obtained from the energy resolution of the light peak, X-rays are used as a reference to determine the energy deposited by the light peak and the noise contribution is determined with a pulse generator. Figure 13 shows the excess noise factor $F$ as a function of gain for different temperatures.

As shown, $F$ does not vary significantly with temperature. All points in Fig. 13 are reasonably fitted to a straight line, so the dependence of $F$ with gain is approximately linear. The results are in agreement with previous measurements, performed at room temperature [10] and liquid nitrogen temperature [39].

Since the variation of F with temperature is not significant, the energy resolution dependence on temperature (Fig. 11) cannot be related to the excess noise factor. Thus, the energy resolution improvement with decreasing temperature should be mainly attributed to dark current.

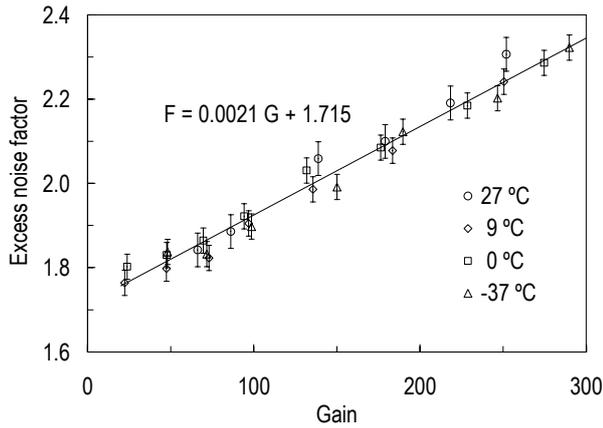

Fig. 13. Excess noise factor as a function of gain for different temperatures. The line is a fit to all data points.

The electronic noise contribution was also determined, by means of a pulse generator, as a function of gain for different temperatures, as shown in Fig. 14. The electronic noise was found to be correlated with dark current: it increases with temperature but does not vary significantly below 0ºC. The electronic noise is the main cause of the energy resolution degradation with temperature.

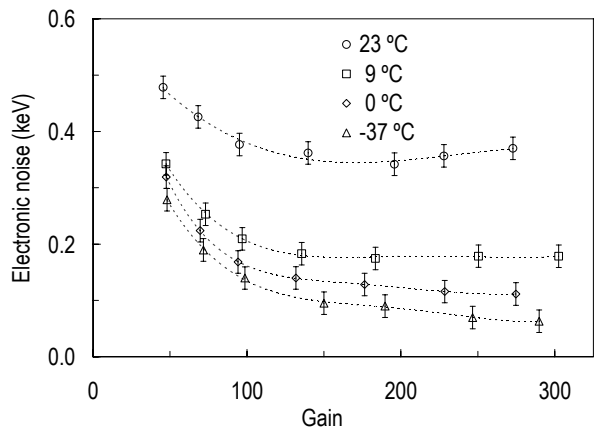

Fig. 14. Electronic noise contribution to the LAAPD energy resolution, normalized to 5.4 keV, as a function of gain, for different temperatures. The lines are used to guide the eyes.

Since the electronic noise is determined by the dark current and the preamplifier characteristics, according to Eq. (9), its variation with gain depends on temperature. For low gains, the noise is mainly determined by the preamplifier, decreasing significantly with gain. For gains above 200, the noise increases slightly at room temperature, while below 0ºC it slightly decreases. The observed behaviour is in good agreement with the predictions of Eq. (9).

The LAAPD intrinsic resolution does not depend on temperature and increases with gain due to the excess noise factor, as results from Eq. (7). So, the main cause of the energy resolution degradation for high gains is the excess noise factor and not the electronic noise contribution. The LAAPD non-uniformity, which also contributes to the energy resolution, is assumed to be gain independent.

### 3.6. Behaviour under intense magnetic fields

The application of LAAPDs as detectors of muonic hydrogen X-rays (1.9 keV) under intense magnetic fields required the investigation of the LAAPD performance for X-ray detection as a function of the magnetic field intensity [18,19]. Figure 15 shows schematically the experimental system used. The LAAPD was placed inside a black box, which protects its surface from the ambient light, and installed together with the preamplifier in a superconducting solenoid which achieves magnetic field intensities up to 5 T. In the depicted geometry, the electric field inside the LAAPD is perpendicular to the magnetic field in the solenoid.

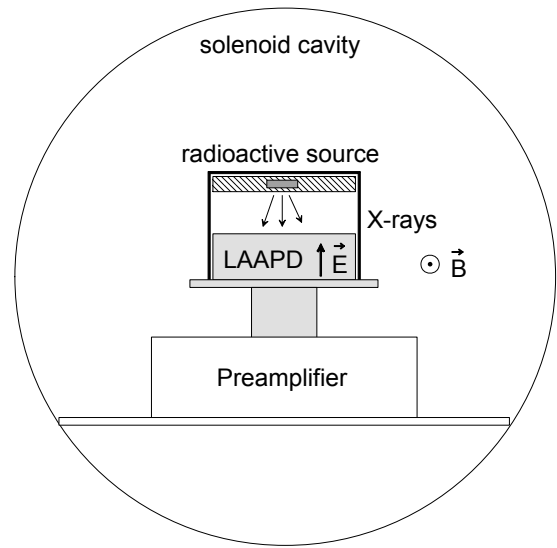

Fig. 15. Schematic of the experimental system used to study the behaviour of LAAPDs for X-ray detection as a function of the magnetic field produced in a superconducting solenoid. The relative orientation of the electric field in the APD and the magnetic fields inside the solenoid is shown.

Another LAAPD was installed inside the solenoid with an axial orientation parallel to the magnetic field. In this case, the trajectory of electrons inside the LAAPD is parallel to the magnetic field and thus is not the affected by the magnetic force.

Both LAAPDs, from API with 16 mm diameter, were polarized with 1800 V. Two radioactive sources, $^{54}$Mn and $^{55}$Fe, were used to irradiate the whole LAAPD area. The LAAPD signals were fed through a charge sensitive preamplifier (Canberra 2004) to a spectroscopy amplifier with 200 ns shaping time constants, being pulse height analyzed in a 1024-channel analyzer.

The LAAPD pulse amplitude and energy resolution were monitored as a function of the magnetic field intensity. Figure 16 shows the relative amplitude and energy resolution for 5.4 and 5.9 keV X-rays, from $^{54}$Mn and $^{55}$Fe sources, being the relative orientation of the electric and magnetic fields perpendicular and parallel, respectively.

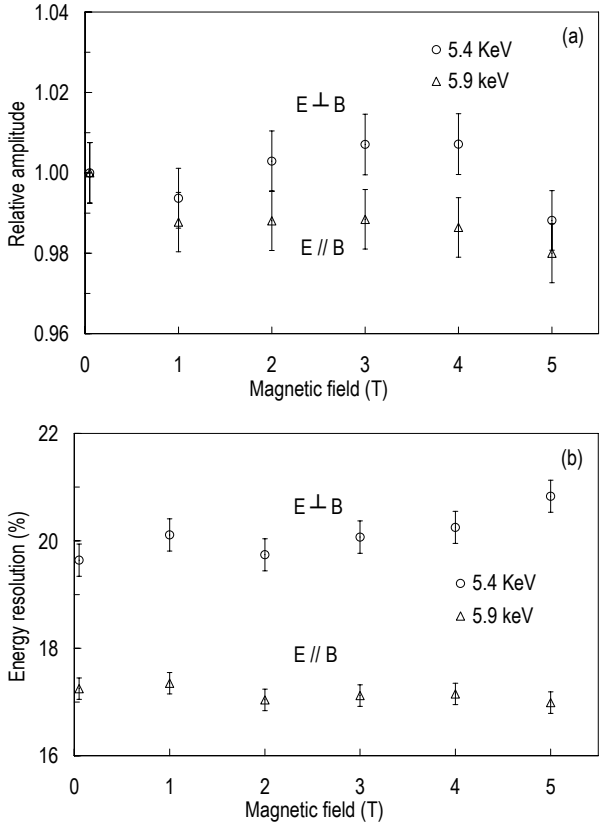

Fig. 16. Relative amplitude (a) and energy resolution (b) of pulses produced by 5.4 and 5.9 keV X-rays in different LAAPDs, with perpendicular and parallel axial orientations relative to the magnetic field, respectively, as a function of the magnetic field intensity.

The relative amplitude variation with magnetic field is lower than 2% for both LAAPDs, being within the experimental errors. The energy resolution variation is also within the experimental errors. There are significant differences on the energy resolution obtained for 5.4 and 5.9 keV X-rays. Besides the energy difference, for the same voltage applied to both LAAPDs the gain and dark current are not equal and thus the energy resolution differs a lot. The LAAPD performance is not significantly affected by the application of magnetic fields up to 5 T, independently on the relative orientation between the electric and magnetic fields.

Figure 17 shows the pulse-height distributions obtained for the LAAPD irradiated with a $^{54}$Mn source. The distributions include the Cr K-lines and the electronic noise tail. As shown, there are no significant differences for 0 and 5 T.

The LAAPD response was also investigated for lower energy X-rays using a sulphur sample irradiated with X-rays from a $^{55}$Fe source. For the signal produced by the 2.3 keV characteristic X-rays, the relative amplitude was found to decrease by less than 3% when the magnetic field increased from 0 to 5 T, while an absolute energy resolution variation of 2% was measured. Both variations are approximately equal to the experimental errors. As a result, LAAPDs can operate under strong magnetic fields up to 5 T without significant performance degradation for X-rays with energy as low as 2.3 keV.

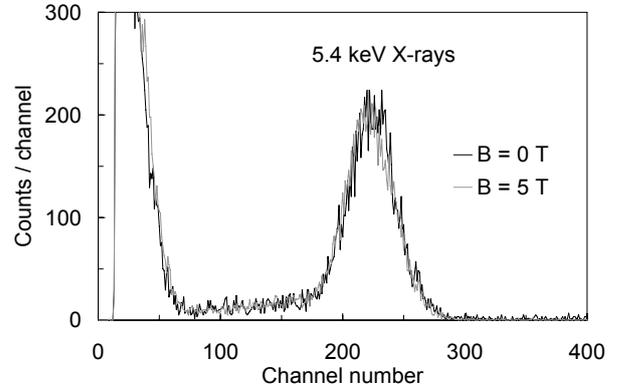

Fig. 17. Pulse-height distributions for 5.4 keV X-rays detected in the LAAPD with axial orientation perpendicular to the magnetic field, for 0 and 5 T.

The time response of LAAPDs was investigated as a function of the magnetic field intensity. Figure 18 shows the time variation of pulses at the preamplifier output produced by 5.4 keV X-rays, for magnetic fields of 0 and 5 T. As shown, the pulse shape does not depend significantly on the magnetic field. The time delay between the two curves is not relevant.

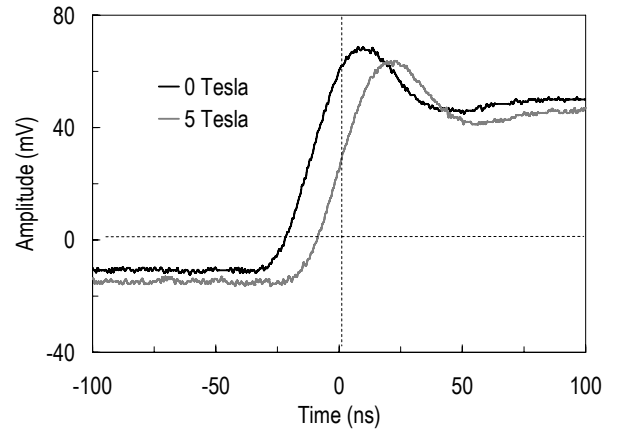

Fig. 18. Pulse-shape at the preamplifier output, resulting from 5.4 keV X-rays detected in the LAAPD, for magnetic fields of 0 and 5 T.

The rise-time was determined as the time interval corresponding to a relative pulse amplitude variation from 10 to 90% in the pulse-shape distributions (Fig. 18). Pulse rise-times of about 25 ns were measured at the preamplifier output for 5.4 keV X-rays. The rise-time variation with magnetic field is not significant. The same trend was observed for 2.3 keV X-rays.

The LAAPD time resolution was determined from the coincidence between 5.4 keV X-rays and 835 keV γ-rays from a $^{54}$Mn source. X-rays are detected by the LAAPD and γ-rays by a NaI(Tl) scintillator readout by a PMT, as shown in Fig. 19. The time-zero is defined by each γ-ray that reach the scintillator. The time resolution is determined from the time spectrum of events detected in coincidence by both detectors.

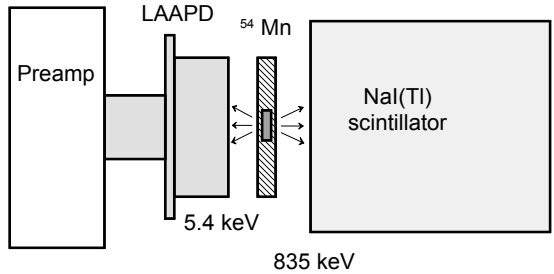

Fig. 19. Experimental system used to measure the LAAPD time resolution. The $^{54}$Mn radioactive source emits 5.4 keV X-rays and 835 keV γ-rays in time coincidence, detected respectively by the LAAPD and a NaI(Tl) scintillator readout by a PMT.

Figure 20 shows a typical time spectrum of signals obtained in coincidence in both detectors. The LAAPD was polarized with 1800 V. The FWHM of the peak in the time distribution defines the time resolution of the system composed of the two detectors. The time resolution measurement was repeated and values between 10 and 12 ns were obtained. Since NaI(Tl) detectors present time resolutions of 3-5 ns and the total time resolution is the quadratic sum of the time resolution of each detector, the LAAPD time resolution is found to be approximately 10 ns.

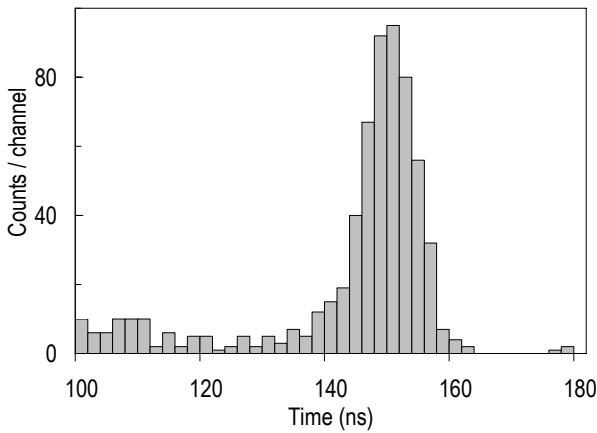

Fig. 20. Time spectrum of coincidences between signals produced in the LAAPD and the NaI(Tl) scintillator by 5.4 keV X-rays and 835 keV γ-rays, respectively. The channel width is 2 ns.

Time resolution measurements were not made as a function of the magnetic field intensity because the NaI(Tl) detector is mechanically sensitive to the field. However, since the amplitude and rise-time do not vary significantly with magnetic field, there are no reasons to expect significant variations of the time resolution.

## 4. VUV-LIGHT DETECTION

Avalanche photodiodes compete favourably with photomultiplier tubes in photon detection applications due mainly to their high quantum efficiency. Windowless VUV-sensitive APDs have been investigated as photosensors of scintillation light produced in noble gases [7,8] and liquids [40,41] for X-ray and γ-ray spectroscopy applications. The application of LAAPDs as VUV photosensors in GPSCs has shown that they can replace PMTs with advantages, including their compact structure, low power consumption, wide dynamic range and high quantum efficiency, covering a wider spectral range. Thus, they provide a more efficient conversion of the scintillation light into charge carriers.

The response characteristics of a LAAPD from API (Deep-UV 500 windowless series) to the scintillation VUV light produced in argon and xenon were investigated at room temperature. This type of APDs is particularly sensitive in the VUV spectral region down to 120 nm. The emission spectra for argon and xenon electroluminescence is a narrow continuum peaking at about 128 and 172 nm, respectively, with 10 nm FWHM for both cases [42], and corresponds to the lower limit of the LAAPD spectral response. For 128 and 172 nm VUV light from the argon and xenon scintillation, the effective quantum efficiency, here defined as the average number of free electrons produced in the APD per incident photon is 0.5 and 1.05, respectively, corresponding to a spectral sensitivity of about 50 and 150 mA/W [43].

An important characterization of LAAPDs used as photosensors is the determination of the number of electron-hole pairs produced per unit of absorbed energy. This enables a quantitative analysis of the number of photons emitted by the luminous source, the noise sources contributing to the energy resolution and the LAAPD quantum efficiency. The number of electron-hole pairs is often determined by comparing the amplitude of the pulses produced by direct absorption of X-rays and by scintillation light. The comparison is valid assuming strict linearity between the initial number of electron-hole pairs and the resulting pulse amplitude throughout the LAAPD range of gains.

Non-linearity is observed at high current densities and attributed to space-charge effects, reduced localized electric fields and heating in the avalanche region, resulting from the local absorption of X-rays [15]. As a consequence, the suitability of using X-rays to determine the number of charge carriers produced in light measurements is somewhat affected by this non-linear behavior.

Significant non-linearity between X-rays and visible light was observed in different types of APDs [3,14,15,26]. LAAPDs produced by API were reported to exhibit reduced non-linearity [4,26]. The light sources used in these studies are LEDs with peak emission in the 400-600 nm range, excluding the deep and vacuum ultraviolet regions.

In addition to the non-linearity between X-rays and light detection, other LAAPD response characteristics to VUV light are different in comparison with visible-light detection, such as the sensitivity to intense magnetic fields and the gain variation with temperature. These different behaviors for visible and VUV light may be related to the photon penetration depth, which is about 1 μm for 520 nm visible photons and about 5 nm for 172 nm VUV photons (xenon scintillation) [29].

The gain non-linearity, LAAPD sensitivity to intense magnetic fields and gain variation with temperature studies were extended to the detection of VUV light produced in argon and xenon [18,24,25].

### 4.1. Performance characteristics for VUV-light

A LAAPD was placed inside a GPSC as a photosensor, as a substitute for the PMT. The detector endowed with the LAAPD, from API (16 mm diameter), is shown in Fig. 21. The GPSC has a 2.5 cm deep absorption or drift region and a 0.8 cm deep scintillation region. The entrance window (6 μm thick Melinex with 2 mm diameter) and grid G1 are operated with negative voltages, while grid G2 is maintained at ground, in order to create appropriate electric fields in the drift and scintillation regions. The LAAPD is supplied by a positive voltage, which determines its gain. The detector was filled with gas, either pure argon at 1140 Torr or pure xenon at 825 Torr, continuously purified by convection through non-evaporable getters (SAES St707).

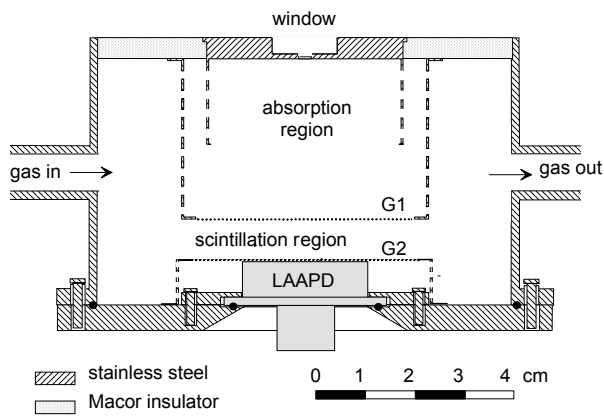

Fig. 21. Schematic of the GPSC instrumented with a LAAPD as VUV photosensor.

Detailed theory and operation of GPSCs are described in Ref. [44-46]. X-rays entering the GPSC window are mostly absorbed in the drift region, producing a cloud of primary electrons that drift toward the scintillation region by influence of a weak electric field (between the window and grid G1). In the scintillation region, each primary electron is accelerated by a stronger electric field (between grids G1 and G2), exciting the gas, without further ionization, and producing a VUV-light pulse (called secondary scintillation or electroluminescence) with about 1 μs duration. Following the incidence of VUV photons on the LAAPD, charge multiplication will take place within the LAAPD volume, originating the final pulse. The operational characteristics of the GPSC with LAAPD are described in Refs. [7] and [8] for xenon and argon filling, respectively.

In order to study the LAAPD response to the VUV light from argon and xenon electroluminescence, the GPSC of Fig. 21 was irradiated with a $^{55}$Fe radioactive source. LAAPD pulses were fed through a low-noise 1.5 V/pC charge-sensitive preamplifier (Canberra 2006) to a linear amplifier (HP 5582A) with 2 μs shaping time constants. Pulse-height analysis was performed with a 1024-channel analyzer (Nucleus PCA-II).

Figure 22 shows typical pulse-height distributions for argon (a) and xenon (b). In (a), only 5.9 keV X-rays from the Mn $K_\alpha$ line interact in the gas because the $K_\beta$ line was efficiently absorbed by a chromium foil. Under the experimental conditions, each 5.9 keV X-ray absorbed in the GPSC drift region produces an average number of $7.5\times10^4$ VUV photons in argon [8] and $1.2\times10^5$ VUV photons in xenon [7]. Taking into account that the average relative solid angle subtended by the LAAPD is approximately 0.2 [47], about $1.5\times10^4$ VUV photons from argon scintillation hit the photodiode active area [8]. For the xenon scintillation, about $2.5\times10^4$ VUV photons hit the LAAPD.

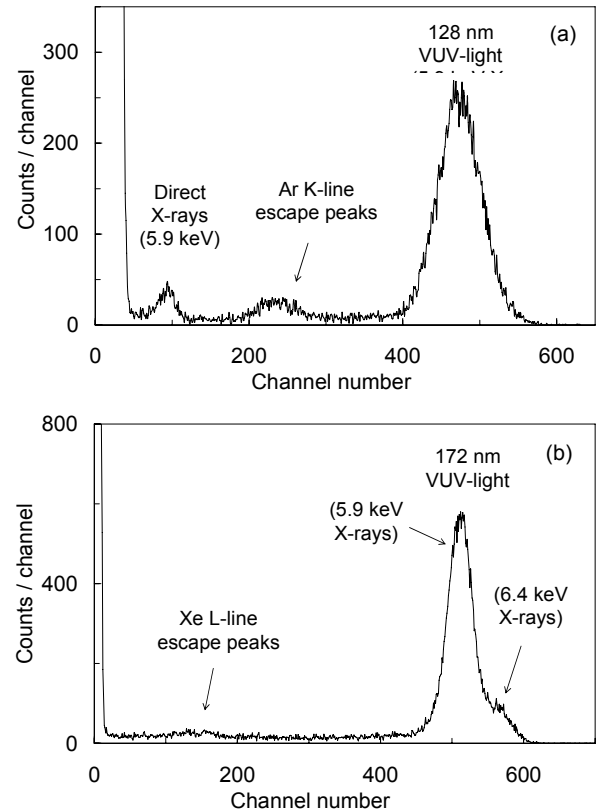

Fig. 22. Typical pulse-height distributions obtained with the detector of Fig. 21, irradiated with a $^{55}$Fe radioactive source and filled with argon (a) and xenon (b). In (a), a chromium filter was used to absorb the Mn $K_\beta$ line from the source. The LAAPD was operated at gains of about 40 and 60 in (a) and (b), respectively.

The main features of the pulse-height distributions include the scintillation peaks resulting from the full absorption of X-rays in the gas and subsequent gas fluorescence escape peaks, as well as the electronic noise tail. In (a), there is an additional peak resulting from direct absorption of 5.9 keV X-rays in the LAAPD. Approximately 10% of the 5.9 keV X-rays are transmitted through the 3.3 cm depth of argon. In (b), there is no peak resulting from transmitted X-rays because the X-ray absorption in xenon is practically 100% and so X-rays do not reach the LAAPD.

The peak resulting from the direct interaction of 5.9 keV X-rays in the LAAPD, for the argon case, can be used to determine the number of charge carriers produced by the VUV-light pulse. As seen in Fig. 22 (a), the amount of energy deposited in silicon by the argon scintillation pulse is similar to what would be deposited by 30 keV X-rays directly absorbed in the LAAPD. The number of primary electrons produced in the LAAPD is about $8 \times 10^3$, which is consistent with the LAAPD quantum efficiency of about 50% for argon scintillation [48].

The LAAPD response to VUV light was investigated as a function of bias voltage for argon and xenon scintillation. The absolute gain was obtained by normalizing the scintillation pulse amplitude to the manufacturer specification for low gains (gain 13.8 at 1577 V). Figure 23 shows the minimum number of detectable photons (MDP) for argon and xenon, defined as the number of photons that would produce a signal in the LAAPD with an amplitude equivalent to the onset of the electronic noise tail.

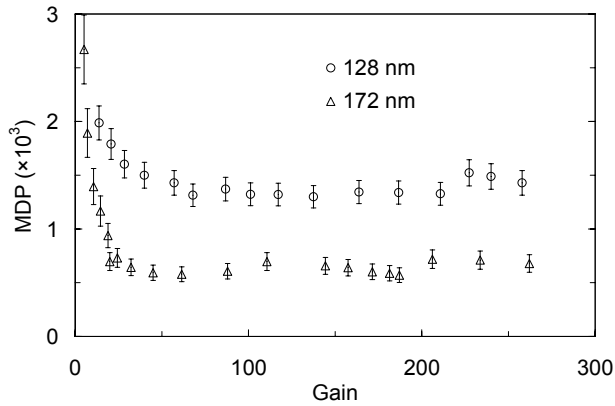

Fig. 23. Minimum number of detectable photons (MDP) for argon and xenon scintillation (around 128 and 172 nm, respectively) as a function of the LAAPD gain.

The MDP variation with gain shows a similar trend for both argon and xenon cases. For low gains, the MDP decreases significantly as the gain drops, stabilizing in 1300 and 600 photons for gains above 60 and 40, respectively for 128 and 172 nm VUV-light pulses.

As expected, the MDP for 172 nm VUV light is about half of the one for 128 nm VUV light, reflecting the ratio between the quantum efficiencies of the LAAPD for those wavelengths, about 105 and 50%, respectively [48]. The obtained MDP can be reduced by cooling the LAAPD due to the reduction of dark current and the electronic noise level, as it will be shown later.

The statistical fluctuations associated to VUV-light detection in the LAAPD can be estimated from the measured energy resolution for the scintillation peaks resulting from the full absorption of 5.9 keV X-rays in the gas. The intrinsic energy resolution of the GPSC is determined by the statistical fluctuations associated to the primary ionization processes in the gas, the production of VUV scintillation photons and the photosensor contribution ($\Delta E$). The statistical fluctuations associated to the scintillation processes in the gas are negligible when compared to the other factors. Thus, the GPSC energy resolution, R, for incident X-rays with energy Ex, is given by [46]:

$$R = 2.355 \sqrt{\frac{fw}{E_x} + \left(\frac{\Delta E}{E}\right)^2} \qquad (13)$$

where $f$ is the Fano factor of the gas, $w$ is the average energy required to create a primary electron in the gas and $E$ is the energy deposited by VUV scintillation light in the photosensor. In the present case (for $E_x = 5.9 \text{ keV}$), $w$ is 26.4 eV for argon [30] and 22.4 eV for xenon [49]. The Fano factor is 0.30 for argon and 0.17 for xenon [50].

The statistical fluctuations associated to the VUV photon detection were derived from Eq. (13). Figure 24 shows the relative standard deviation ($\Delta E/E$) associated to the detection of $1.5 \times 10^4$ argon VUV photons and $2.5 \times 10^4$ xenon VUV photons as a function of the LAAPD gain. The relative uncertainty decreases rapidly with the onset of gain and stabilizes for gains above 30, reaching values of 3.9 and 2.2% for argon and xenon scintillation, respectively.

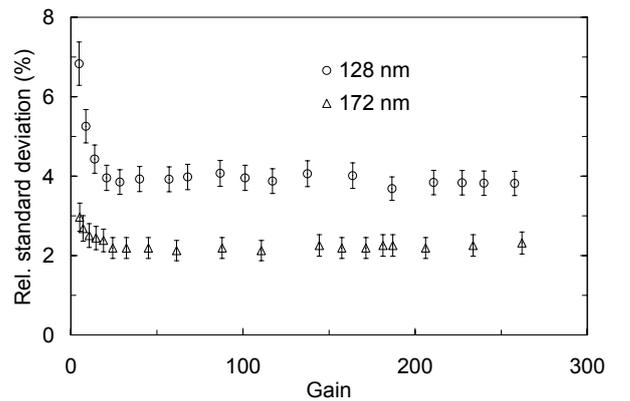

Fig. 24. Relative standard deviation associated to the detection of $1.5 \times 10^4$ argon VUV photons (128 nm) and $2.5 \times 10^4$ xenon VUV photons (172 nm), as a function of the LAAPD gain.

Figure 24 shows that gains as low as 30 are sufficient to achieve the optimum performance. However, the best performance characteristics for the detection of 128 and 172 nm VUV photons are achieved for gains of about 60 and 40, respectively (Fig. 23). For lower light levels, higher gains will be needed to pull the light pulse amplitude above the noise level and achieve the best possible performance.

The LAAPD is a detector suitable for the detection of VUV-light photons with wavelength down to about 120 nm. The use of X-rays as reference is a straightforward method to determine the number of the VUV photons in the light pulse. The LAAPD is not suitable for single photon detection and VUV photon spectrometry, but it can be applied to synchrotron radiation as a VUV photon detector and to other areas of optics. The minimum number of detectable photons and the statistical fluctuations associated to VUV detection do not depend directly on the photon wavelength, but rather on the number of charge carriers produced by the light pulse in the LAAPD, in opposition to non-linearity effects and behaviour under intense magnetic fields, described next.

*4.2. Non-linearity between X-ray and VUV-light gains*

The gain non-linearity study requires the comparison between the amplitude of pulses produced by X-rays directly absorbed in the LAAPD and the one for signals resulting from the scintillation light produced in the gas. Using the GPSC shown in Fig. 21, the study is possible only for argon because a fraction of X-rays is transmitted through the gas, being directly absorbed in the photodiode. The resulting X-ray signals are visible in the pulse-height distribution of Fig. 22 (a). Since the X-ray transmission in xenon is much lower than that in argon, the non-linearity study for xenon scintillation was performed with a different GPSC, made without drift region, also instrumented with a LAAPD (Fig. 25). This driftless GPSC has a 1.1 cm deep scintillation region and is operated at 1140 Torr of xenon continuously purified through getters (SAES St172). A negative voltage is supplied to the entrance window (12.5 μm thick Mylar with 10 mm diameter), while the grid is maintained at 0 V, establishing the electric field required to produce electroluminescence.

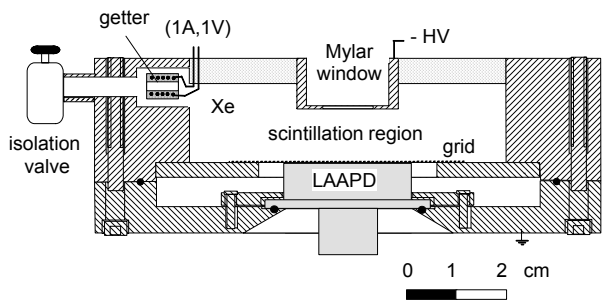

Fig. 25. Schematic of the GPSC driftless prototype with a LAAPD as photosensor filled with xenon at 1140 Torr.

The operation of the driftless GPSC is similar to the one of the prototype previously described with the exception that X-rays are now absorbed in the scintillation region. Thus, the total number of scintillation photons produced by primary electrons accelerated through the electric field depends on how deep the X-ray absorption occurs in the scintillation region. As a result, the driftless design results in worse energy resolution. Nevertheless, about 0.2% of X-rays are transmitted through the 1.1 cm thickness of gas at 1140 Torr, interacting directly in the LAAPD.

Figure 26 shows a typical pulse-height distribution for 5.9 keV X-rays incident on the detector shown in Fig. 25. Since the detector does not have a drift region, the scintillation peak resulting from 5.9 keV X-ray absorptions in the gas is asymmetric and presents larger energy resolution than the one of Fig. 22 (b). In addition to the scintillation peaks, peaks resulting from direct X-ray absorptions in the LAAPD are also visible. These include 5.9 keV X-rays from the $^{55}$Fe source and xenon fluorescence X-rays (4.1 and 4.4 keV from $L_\alpha$ and $L_\beta$ lines).

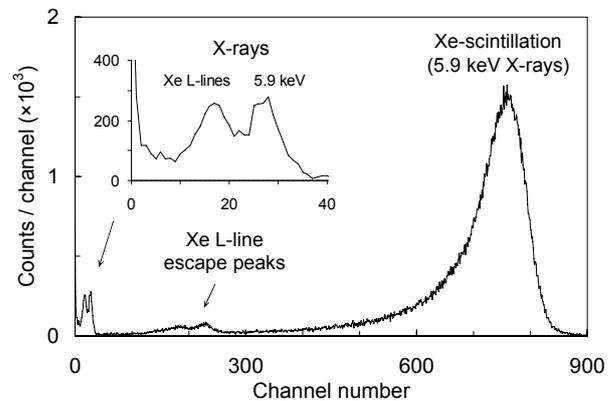

Fig. 26. Pulse-height distribution for 5.9 keV X-rays interacting in the driftless GPSC, filled with xenon at 1140 Torr, operated at a reduced electric field of 5.5 V cm$^{-1}$ Torr$^{-1}$ in the scintillation region and a gain 100 in the LAAPD.

The comparison between the pulse amplitude for scintillation photons and X-rays absorbed in the LAAPD was made as a function of the LAAPD bias voltage. The absolute gain was determined from the amplitude measurements for VUV-light pulses, normalizing to the LAAPD specification at low gains. The gain for X-rays is assumed to be equal to the gain for VUV-light. Figure 27 shows the pulse amplitude ratio (relative gain) between 5.9 keV X-rays and scintillation light produced in argon and xenon, as a function of the VUV-light gain. The variation is approximately linear and, for a gain 200, reaches about -7 and -10% for xenon and argon, respectively.

The non-linearity study was extended to X-rays with higher energy. For 22.1 keV X-rays (from a $^{109}$Cd source) interacting in the xenon GPSC, the gain ratio between X-rays and xenon VUV light presents significantly higher variations than for the case of 5.9 keV X-rays, reaching about -13% for a gain 200. This result confirms the non-linearity between X-rays with different energy, attributed to space charge effects, and is consistent with that presented in Fig. 7.

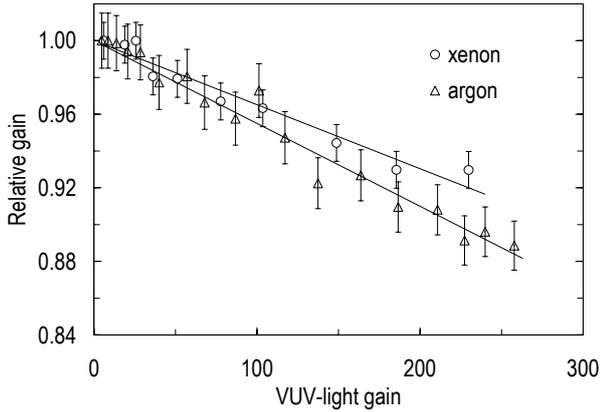

Fig. 27. LAAPD relative gain between 5.9 keV X-rays and VUV light from argon and xenon scintillation, as a function of the light gain. Linear fits to the data are also shown.

In addition to space charge effects resulting from the point-like nature of X-ray interactions, our results suggest a dependence of the non-linearity on the light wavelength. The X-ray to light gain non-linearity for 128 nm VUV photons is higher than the one obtained for 172 nm VUV photons and both are higher than that reported for visible light (~600 nm) [26]. The non-linearity must depend on the penetration depth in silicon of each type of light. For both VUV and visible light, photons are absorbed in the drift region of the LAAPD, where the electric field is very weak and the effect of capture of charges is more significant. Since the absorption is much more superficial for VUV light (~5 nm), capture is greater in this case but decreases with gain due to the electric field increase. Thus, the higher LAAPD voltage helps to collect charges produced near the entrance surface for argon, whereas this effect is smaller for xenon and probably negligible for visible light since the penetration depth increases.

Accurate determination of the number of charge carriers produced in the LAAPD by VUV light, especially at high gains, using X-rays as a reference have to take into account the LAAPD non-linearity.

*4.3. Temperature dependence*

In order to investigate the LAAPD response to VUV-light as a function of temperature, a LAAPD with an integrated Peltier cell, from API (571 cooled head series), was incorporated in a GPSC filled with pure xenon at 825 Torr. The LAAPD is similar to the previous ones. The system includes a thermistor that provides temperature measurements within ± 0.1 ºC. The detector was similar to the one shown in Fig. 21. The absorption and scintillation regions were 3.1 and 0.9 cm thick, respectively.

To produce a fixed amount of VUV photons, the GPSC was irradiated with 5.9 keV X-rays from a $^{55}$Fe source. In the experimental conditions, each X-ray absorbed in the drift region originates about $1.35 \times 10^5$ VUV photons. Taking into account that the relative solid angle subtended by the LAAPD is 0.2 [7], about $2.7 \times 10^4$ VUV photons are detected in the LAAPD per X-ray interaction in the gas.

The pulse-height distributions for LAAPD signals produced by 5.9 keV X-rays incident on the GPSC were obtained for different operation temperatures and LAAPD bias voltages. The pulse-height distributions are similar to that of Fig. 22 (b), with the exception that the Mn K$_\beta$ line (6.4 keV X-rays) from the $^{55}$Fe source is not present because a chromium filter was used.

The gain was determined from amplitude measurements of VUV-light pulses as a function of the LAAPD bias voltage for different temperatures. The gain increases as the temperature decreases. The maximum achievable gain increases from 300 to above 700 as the temperature decreases from 25 to -5 ºC [25].

Figure 28 shows the LAAPD gain for xenon scintillation light as a function of temperature, for different bias voltages. Exponential fits to the data were made. For each voltage, the relative gain variation is almost constant and increases from -2.7% to -5.6% per ºC as the voltage increases from 1633 to 1826 V. For gains around 200 and at room temperature, the results show relative gain variations that are almost a factor two higher than the -3% per ºC reported for visible light [27].

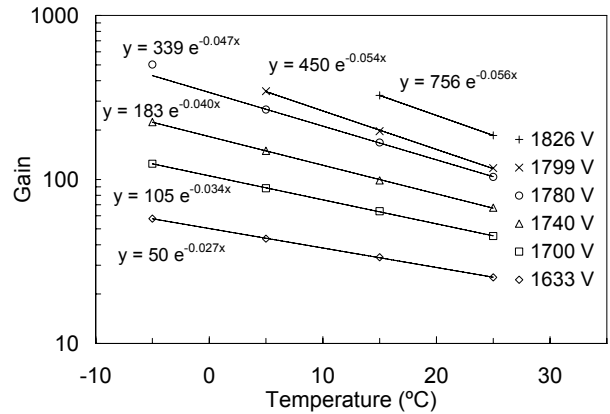

Fig. 28. LAAPD gain for xenon scintillation light (~172 nm) as a function of temperature, for different bias voltages. The lines are exponential fits to the points.

The minimum number of detectable VUV photons (MDP) and the energy resolution for xenon scintillation ($2.7 \times 10^4$ VUV photons) were also determined as a function of gain for different temperatures. The statistical fluctuations associated to VUV-light detection were then estimated from Eq. (13). Figure 29 shows the relative standard deviation associated to these fluctuations (a) and the MDP (b) as a function of LAAPD gain, for different temperatures. The MDP tends to stabilize for gains above 100 and improves as the LAAPD temperature decreases. It seems that the LAAPD operation at temperatures below 5 ºC does not improve its performance in VUV-light detection, except for higher gains. However, since the LAAPD cooling system by Peltier cell was designed to operate between 0 and 40 ºC, the results for temperatures below 0 ºC are not conclusive. Even so, the LAAPD optimum performance characteristics are achieved for gains around 100 and do not depend significantly on temperature.

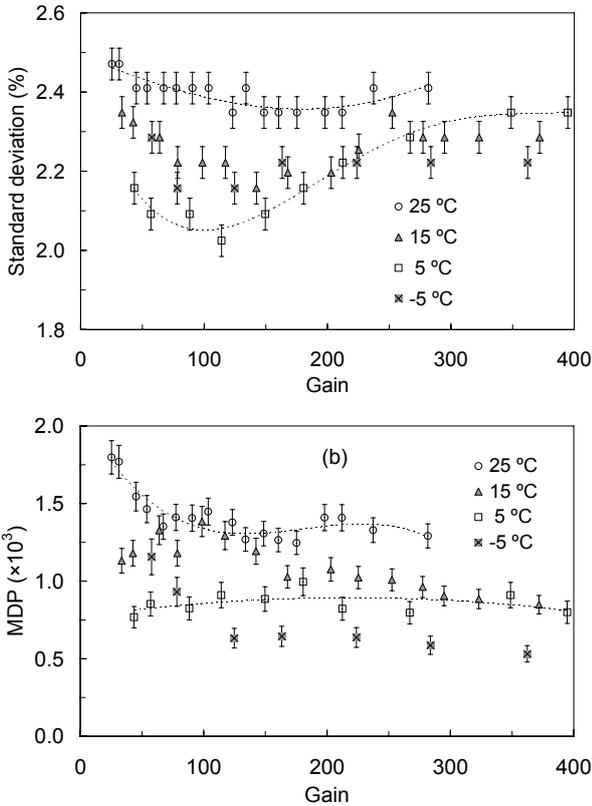

Fig. 29. Relative standard deviation associated to the detection of $2.7 \times 10^4$ VUV photons (a) and minimum number of detectable VUV photons (b) as a function of the LAAPD gain, for different temperatures. The doted lines shown for 25 and 5ºC are used to guide the eyes.

*4.4. Behaviour under intense magnetic fields*

The insensitivity of avalanche photodiodes to magnetic fields has often been referred to in the literature [2,28], but detailed experimental results are scarce. In addition, most of the studies were carried out for visible light and for magnetic field intensities of a few Tesla. Because some LAAPD characteristics are different for visible and VUV-light detection, the behaviour of LAAPDs under magnetic fields up to 5 T was investigated for the two types of radiation.

The study of the LAAPD response to VUV and visible light as a function of magnetic field was performed in the superconducting solenoid described in section 3.6. For VUV-light detection, the driftless GPSC of Fig. 25, filled with xenon at 1140 Torr, was used. A collimated X-ray beam (2 mm diameter) from a $^{55}$Fe radioactive source was made to interact in the xenon producing VUV light. For visible-light detection, a 1 cm$^3$ cubic CsI (Tl) crystal was irradiated with 835 keV γ-rays emitted by a $^{54}$Mn radioactive source. The scintillation light (~520 nm) produced by each γ-ray interaction was detected by a LAAPD. For both cases, the LAAPD was operated at 1800 V and the amount of scintillation light collected in the LAAPD was kept constant. Shaping time constants of 1 and 2 μs were used in the linear amplifier for visible and VUV light measurements, respectively.

Figure 30 shows the pulse-height distributions obtained for visible light (a) and VUV light (b), for magnetic fields of 0 and 5 T. In (a), the amplitude distribution includes the peak corresponding to full energy absorption of γ-rays in the crystal, the Compton continuum and the electronic noise tail. In (b), spectral features include the peaks corresponding to the absorption of Mn K-line X-rays in the xenon GPSC, the gas fluorescence escape peaks and the electronic noise tail. As shown, for visible-light detection there is no significant degradation of the pulse-height distribution when the magnetic field varies from 0 to 5 T, while for the VUV-light detection an amplitude reduction larger than 20% is observed.

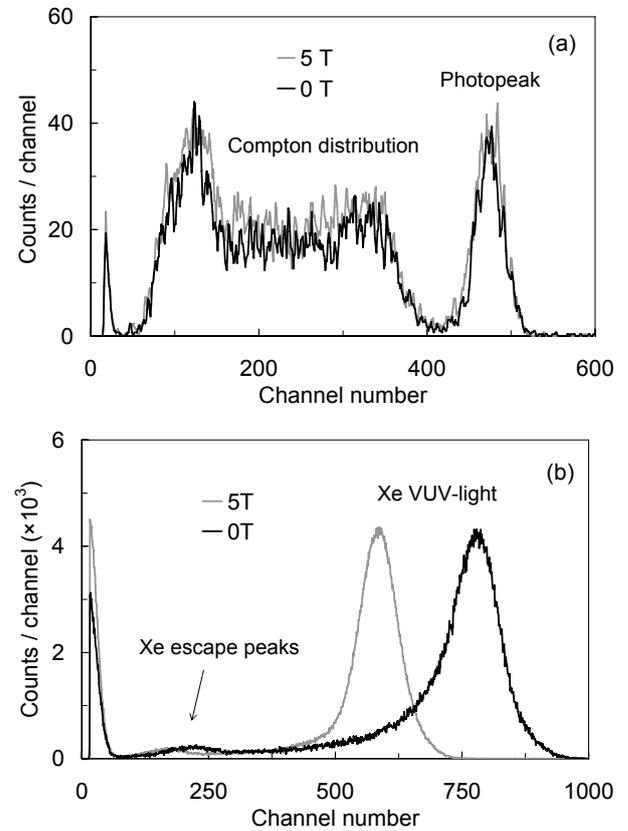

Fig. 30. Pulse-height distributions for the scintillation light detected in a LAAPD: (a) resulting from 835 keV γ-rays interacting in a CsI(Tl) crystal; (b) resulting from 5.9 keV X-rays absorbed in a xenon GPSC.

Figure 31 shows the LAAPD relative pulse amplitude (a) and energy resolution (b) as a function of the magnetic field, for CsI(Tl) scintillation visible light and xenon scintillation VUV light. For visible-light pulses, amplitude variations are within the experimental errors (± 1%) and no significant energy resolution variations are observed. For VUV-light pulses, the relative amplitude decreases gradually as the magnetic field is applied, reaching a 24% variation at 5 T, while the energy resolution increases from 13% to 15%. Since the effect of magnetic field on the GPSC scintillation and possible variations of the solid angle subtended by the LAAPD due to the Lorentz angle are negligible [51], the noticeable influence of the magnetic field in VUV-light detection has its origin in the LAAPD. Since VUV photons interact within the first few atomic layers of silicon, where the electric field is weaker, the magnetic field influences the produced photoelectrons and the diffusion of secondary electrons, originating partial charge loss to the dead layer at the LAAPD entrance.

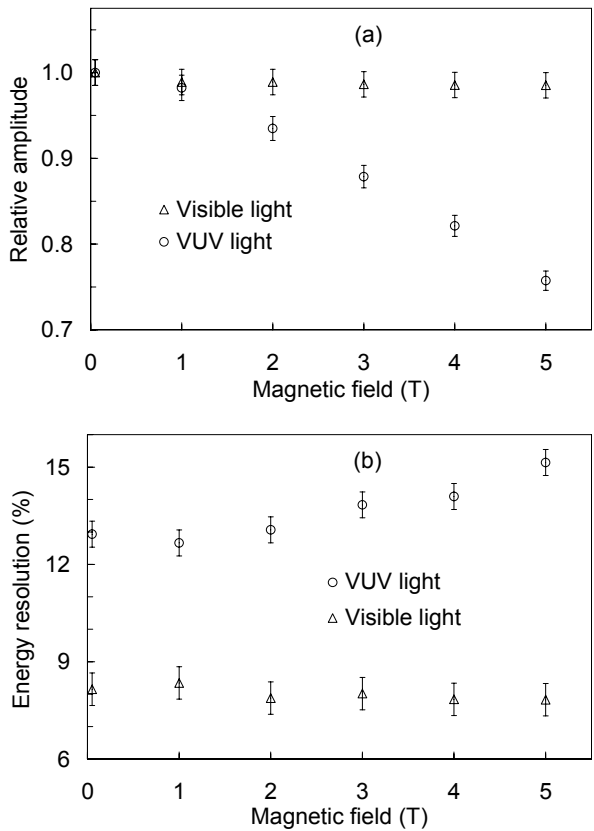

Fig. 31. Relative pulse amplitude (a) and energy resolution (b) as a function of the magnetic field intensity, for visible light and xenon VUV light.

## 5. X-RAY SPECTROMETRY APPLICATIONS

The LAAPD can be used for X-ray spectrometry applications either as VUV photosensor in a GPSC or as direct X-ray detector. The application of LAAPDs in GPSCs has been investigated [7,8] and it was proved that LAAPDs have advantages compared to traditional PMTs.

The performance of the GPSC is optimum using xenon as the scintillation gas. Moreover, the LAAPD quantum efficiency is higher for the xenon scintillation light when compared to other gases. An energy resolution of 7.8 % was obtained for 5.9 keV X-rays interacting in the GPSC. This value is similar to the one obtained using a PMT with an area 10 times larger than the LAAPD. For X-rays with lower energy, the resolution is better with the PMT due to the higher gain and consequently higher signal-to-noise ratio. However, above 6 keV, the energy resolution is better with the LAAPD.

In this chapter, different applications of LAAPDs as direct X-ray detectors are described. The application to X-ray fluorescence analysis was investigated at room and lower temperatures. The application of digital pulse rise-time discrimination techniques to LAAPDs was investigated in order to improve their performance for low-energy X-ray spectrometry.

### 5.1. Energy linearity and X-ray fluorescence analysis at room temperature

The application of avalanche photodiodes to X-ray fluorescence analysis was investigated at room temperature by detecting X-rays resulting from several samples irradiated with $^{55}$Fe and $^{109}$Cd radioactive sources.

The LAAPD energy linearity was investigated in the X-ray energy range between 1.7 and 25 keV using the fluorescence radiation induced in single-element samples (Si, S, Cl, Ca, Ti, Cr, Fe, Ni, Zn, As, Se, Rb, Nb) and using direct X-rays emitted by the radioactive sources. The X-ray beams incident on the LAAPD were collimated using lead collimators with 1 and 5 mm diameters, respectively for direct irradiation and fluorescence measurements.

For the energy linearity study, the LAAPD was operated in the optimum gain region (around 50) with total counting rates of $10^3$ to $10^4$/s. The peak centroid and FWHM in the different pulse-height distributions were determined. Figure 32 shows the pulse amplitude (peak centroid) and the energy resolution as a function of X-ray energy. Good linearity is observed throughout the energy range for the optimum gain. For higher gains, non-linearity effects will take place.

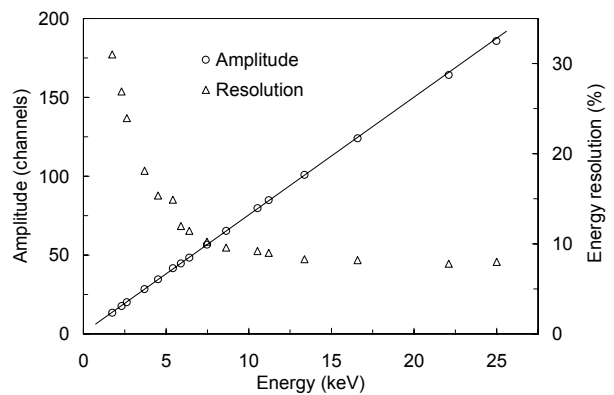

Fig. 32. LAAPD pulse amplitude (peak centroid) and energy resolution as a function of X-ray energy. A linear fit to the amplitude measurements is shown. The experimental uncertainties fall within the symbol size.

The energy resolution does not scale as $E^{-1/2}$ ($E$ being the X-ray energy), especially for higher energies, as it happens in proportional counters. This behaviour is due to peak distortion resulting from an increasing number of X-ray interactions in the multiplication region for higher energies. Nevertheless, LAAPDs may achieve better energy resolution compared to typical proportional counters, especially at lower energies and for devices with low dark current.

X-ray fluorescence spectra were measured for thick samples of gypsum (CaSO4) and SAES St707 getters (70% Zr, 5.4% Fe and 24.6% V), irradiated with 55Fe and 109Cd sources, respectively. The resulting pulse-height distributions are shown in Fig. 33 (a) and (b), respectively, and include the characteristic K-lines of the sample elements as well as the backscattered radiation from the X-ray source (Mn and Ag K-lines for $^{55}$Fe and $^{109}$Cd, respectively). In Fig. 33 (b), Pb L-lines from the source shielding and collimator are also present. The amplitude spectra and energy resolutions are similar to those achieved with conventional proportional counters.

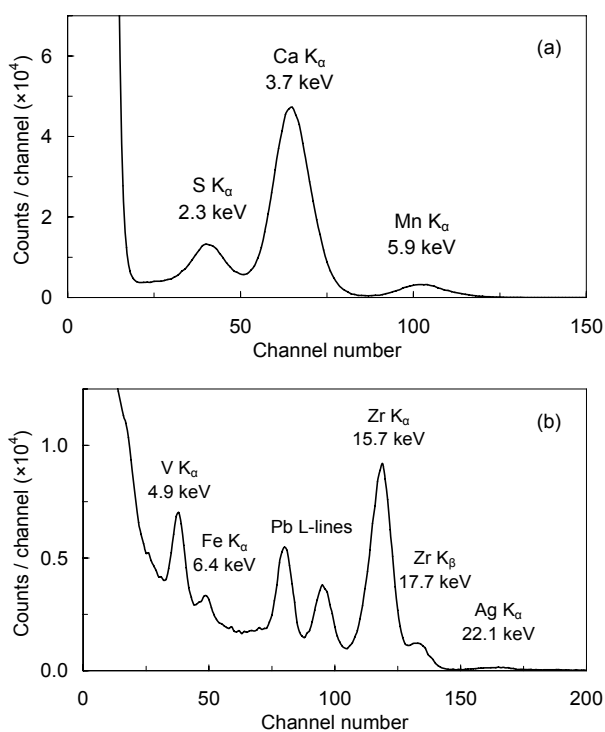

Fig. 33. X-ray fluorescence spectra obtained from two thick samples: (a) gypsum (CaSO4) irradiated with a 55Fe source; (b) SAES St707 getters (70% Zr, 5.4% Fe and 24.6% V) irradiated with a 109Cd source.

The previous results demonstrate the applicability of LAAPDs, working at room temperature, to energy-dispersive X-ray fluorescence analysis, where they can be used to detect X-rays in the 2-25 keV energy range. Compared to proportional counters, the LAAPD is more compact and may provide improved energy resolution for X-ray energies from a few keV up to about 20 keV. The LAAPD cost, limited area, limited detection efficiency for medium and high energy X-rays, as well as the sensitivity to light and temperature, are drawbacks compared to proportional counters. However, the LAAPD windowless feature may be crucial for detection of soft X-rays. Additionally, its superior counting rate capability may be important for applications with high counting rates.

### 5.2. X-ray fluorescence analysis at low temperatures

In order to investigate the applicability of LAAPDs to X-ray fluorescence analysis at different temperatures, the LAAPD energy linearity and energy resolution were determined in the energy range from 3 to 22 keV using the fluorescence radiation induced in single-element samples by $^{55}$Fe and $^{109}$Cd radioactive sources. Figure 34 plots the relative amplitude and energy resolution as a function of the X-ray energy for different temperatures. The LAAPD was operated at a gain 130.

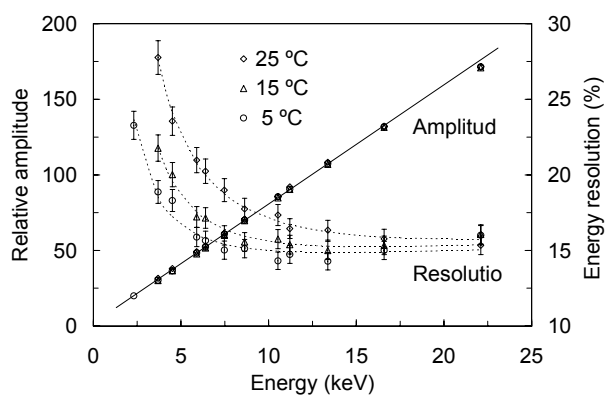

Fig. 34. Relative amplitude and energy resolution for LAAPD X-ray pulses as a function of X-ray energy, for different operation temperatures and a LAAPD gain 130. For amplitude measurements, errors are within the symbol size. The doted lines are used to guide the eyes while the full line is a linear fit to all amplitude measurements below 17 keV.

Good energy linearity is observed for energies below 20 keV for the used gain, independent on temperature. The energy resolution improves with decreasing temperature due to the diminishing dark current. As expected, the energy resolution decreases with the X-ray energy, but tends to stabilize for higher energies due to the larger distortion of X-ray peaks.

Figure 35 shows typical pulse-height distributions obtained for fluorescence X-rays from a sulphur sample irradiated with a $^{55}$Fe source, for different temperatures. The distributions also include the X-ray source Mn-backscattered line. About 240, 280 and 320 gains were used respectively for 25, 15 and 5ºC. The energy spectra are better than those obtained with a conventional proportional counter, especially for lower temperatures. The energy resolution for 2.3 keV X-rays (S K$_\alpha$-line) improves from 30 to 23% as the temperature decreases from 25 to 5ºC and the same happens to the minimum detectable energy, which improves from 1.7 to 1.1 keV.

Figure 36 shows the pulse-height distributions for X-ray fluorescence of a gypsum (CaSO$_4$) sample irradiated with a $^{55}$Fe source, for 26, 2 and -21ºC operation temperatures and a LAAPD gain 100. The distributions include the X-ray fluorescence peaks from calcium and sulphur and the source backscattering peak (K-lines).

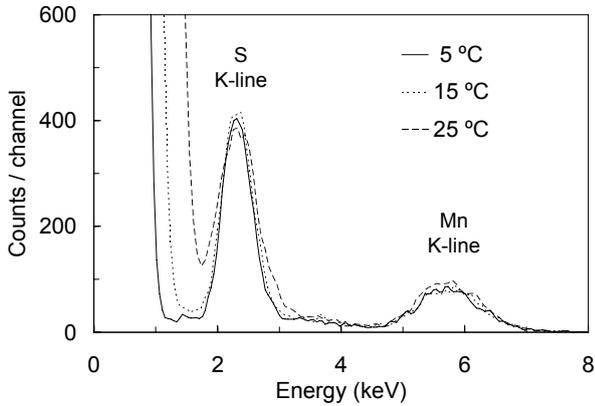

Fig. 35. Pulse-height distributions for fluorescence X-rays from a sulphur sample irradiated with a $^{55}$Fe X-ray source, for different operation temperatures.

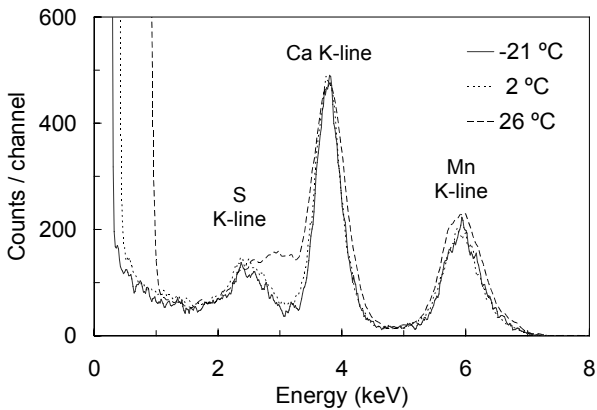

Fig. 36. LAAPD pulse-height distributions for X-ray fluorescence of a CaSO$_4$ sample irradiated with a $^{55}$Fe source, for different temperatures.

From the comparison of Figs. 35 and 36, it is noticeable that the minimum detectable energy is better in Fig. 36. Two different LAAPDs, with different operating gains were used, so electronic noise is different in both cases. The relative background level under the sulphur-line is higher in Fig. 36 since more X-ray lines with higher energies are present. The energy resolution of the S K$_\alpha$-line (2.3 keV) seems to be worse in Fig. 36 but statistics are poorer. Decreasing temperature shows no significant variation of the background level.

The advantages of the LAAPD cooling in its performance are evidenced in the previous energy spectra by both electronic noise reduction and energy resolution improvement for decreasing temperatures. However, for temperatures below 0ºC only a small improvement is observed. The effect of the temperature reduction is more significant for fluorescence X-rays with lower energies, closer to the noise level.

### 5.3. Application of pulse rise-time discrimination to LAAPDs

In some cases, such as X-ray spectrometry, the background created by low-amplitude X-ray signals may limit the application of LAAPDs. One way to improve the LAAPD performance, mainly in the low energy range, may be through the application of pulse discrimination techniques in order to efficiently distinguish fully amplified X-ray pulses from noise, distorted X-ray pulses and pulses resulting from the interaction of charged particles.

The application of rise-time discrimination techniques to LAAPDs was investigated in order to improve their response to X-rays [17]. X-rays absorbed in the undepleted p-region of the LAAPD produce pulses with long time responses as traps may hold electrons for tens to thousands of nanoseconds [28]. Moreover, the time response for events resulting from interactions in the multiplication region is faster than the one for fully amplified events. These different time responses may result in pulse rise-time differences, making possible the discrimination and rejection of such anomalous pulses and, consequently, improved detector performance.

The used method includes signal analogue pre-shaping followed by digital processing in a commercial digitizer operated at its maximum sampling frequency of 125 MHz. The digital pulse-height analyzer and the discrimination method are described in Refs. [17,52,53]. The pulse rise-time is defined as the time required for the pulse amplitude to increase from 10 to 90%.

The digital discrimination method was applied to direct irradiation of a LAAPD with a collimated beam (2 mm diameter) of 5.9 keV X-rays. Figure 37 presents two pulse rise-time distributions using 200 ns shaping time constants and two different trigger thresholds: just below and above the LAAPD noise level.

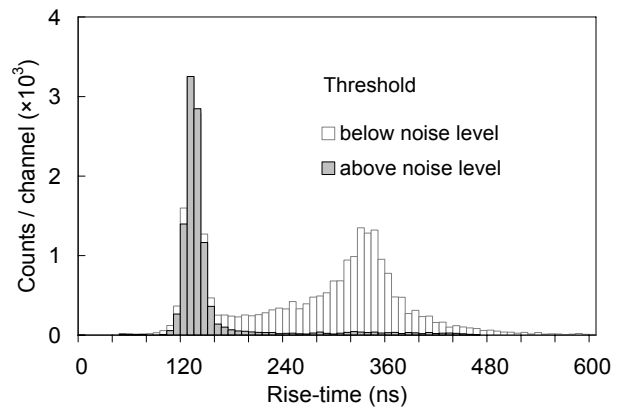

Fig. 37. Distributions of pulse rise-times for 5.9 keV X-rays detected in the LAAPD, using 200 ns shaping time constants in the linear amplifier and two different trigger thresholds in the digital pulse-height analyzer: just below and above the LAAPD noise level. The channel width is 8 ns.

Figure 37 shows that noise pulses have longer rise-times. Lowering the trigger threshold in the digital pulse-height analyzer results in reduced data acquisition throughput due to the time required to analyze noise pulses, during which acquisition is stopped.

Figure 38 presents the pulse-height distributions for 5.9 keV X-rays, including the raw spectrum and two partial distributions for different rise-time intervals (128-160 and 136-152 ns). The trigger threshold was set just above the LAAPD noise level to avoid dead time losses. Each distribution was fitted to a Gaussian function superimposed on a linear function in order to determine the peak centroid, FWHM and area.

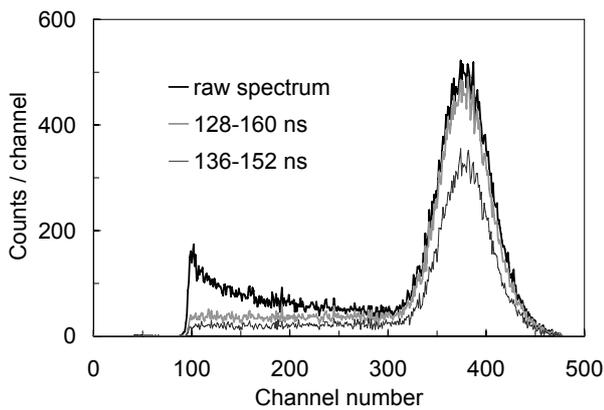

Fig. 38. Pulse-height distributions for 5.9 keV X-rays, including the raw spectrum and two partial distributions for rise-time intervals of 128-160 and 136-152 ns.

The energy resolution and peak-to-background ratio, defined as the Gaussian area divided by the area under the linear function in the peak region, were determined for the distributions of Fig. 38 and for distributions of narrower rise-time intervals. Table 4 shows the characteristics of the 5.9 keV X-ray peak for several partial distributions corresponding to different rise-time windows.

Table 4. Characteristics of the 5.9 keV X-ray peak for relevant partial distributions with different rise-time windows.

| Rise-time window (ns) | Centroid (channel) | Energy resolution (%) | Peak-to-background | Relative area (%) |
|---|---|---|---|---|
| full range | 379.2 | 16.4 | 21.4 | 100 |
| 128-136 | 365.6 | 14.9 | 28.6 | 15 |
| 136-144 | 375.6 | 15.3 | 31.7 | 33 |
| 144-152 | 383.6 | 14.9 | 31.9 | 30 |
| 152-160 | 389.7 | 14.3 | 25.7 | 13 |

For the partial distributions of Fig. 38, the improvement of the energy resolution and peak-to-background ratio is not as significant as for the narrower rise-time intervals of Table 4. In the first case, the energy resolution was found to be 15.5 and 16.0% for rise-time intervals of 136-152 and 128-160 ns, corresponding to 63% and 91% counting efficiency (relative peak area). The peak-to-background ratio was 26 and 30, respectively. For narrower intervals (Table 4), the energy resolution and peak-to-background ratio show some improvements, while the counting rate efficiency decreases. Thus, a balance between counting rate and rise-time discrimination must be found.

In addition, Table 4 shows that the peak centroid increases with the rise-time due to the ballistic deficit, which measures the signal amplitude loss resulting from incomplete signal integration. This happens because some charge is lost through the polarization resistance in series with the detector during the voltage signal creation.

The effect of the ballistic deficit can be somewhat corrected by normalizing the peak centroids of the partial distributions. The centroid of each partial distribution corresponding to the rise-time windows of Table 4 was normalized to the centroid of the raw spectrum. Then, distributions were added, originating a single distribution with rise-times between 128 and 160 ns. The energy resolution for this distribution improves from 16.0% (Fig. 35) to 15.1%, while the peak-to-background ratio is not affected by this ballistic deficit correction.

Figure 38 shows that a relevant feature of the rise-time discrimination is the reduction of low-energy tails associated to X-ray peaks, an important advantage for X-ray spectrometry when several X-ray lines are observed and lower energy peaks superimpose on the background of higher energy lines.

Pulse rise-time distributions were measured for different X-ray energies, with the LAAPD operated at the optimum gain [17]. The rise-time distribution is similar for all tested X-ray energies (between 2 and 11 keV), with an average rise-time of 134 ns.

Figure 39 shows the pulse-height distributions obtained for a silicon sample irradiated with a $^{55}$Fe source. The raw spectrum and two partial distributions, obtained for rise-time windows of 120-192 ns and 120-152 ns, are shown. Each distribution includes three different peaks, corresponding to the Si (1.7 keV), Ca (3.7 keV) and Mn (5.9 keV) fluorescence $K_\alpha$-lines. The calcium line results from plasticine used to hold the radioactive source.

As noise pulses can be efficiently discriminated against X-ray pulses, the effect of rise-time discrimination is better for X-rays of lower energy, whose amplitude is closer to the noise level.

In the raw spectrum of Fig. 39, the Si peak area and background cannot be determined with a reasonable precision since the noise tail is superimposed on the peak region. By fitting the noise to a Gaussian, energy resolution values of 43.6, 26.5 and 17.9 were obtained for Si, Ca and Mn K-lines, respectively. In the partial distributions, the energy resolution for the Si $K_\alpha$-line (1.7 keV) improved to 34.4 and 31.3% for rise-time windows of 120-192 and 120-152 ns, respectively, while the peak-to-background ratio did not change significantly. For these partial distributions, the energy resolution for the Ca $K_\alpha$-line (3.7 keV) slightly improved to 26.0%, while for the Mn $K_\alpha$-line (5.9 keV) no energy resolution variation was noticed.

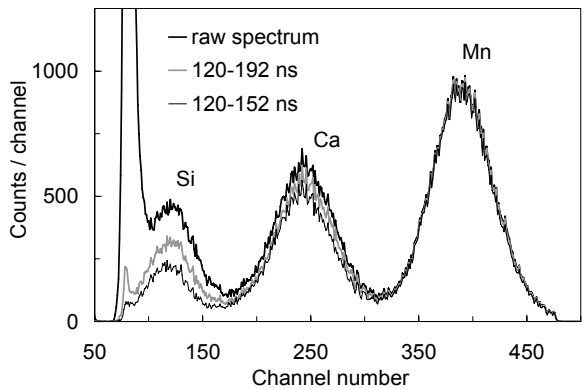

Fig. 39. Pulse-height distributions for fluorescence X-rays (K-lines) from Si, Ca and Mn, obtained from a silicon sample and a plasticine holder irradiated with a $^{55}$Fe source.

Figure 40 shows the raw spectrum and several partial pulse-height distributions spanning the full rise-time range (0-400 ns). As seen, the majority of X-ray events are in the 120-152 ns rise-time interval and noise pulses present longer rise-times, so they can be efficiently discriminated against X-ray pulses.

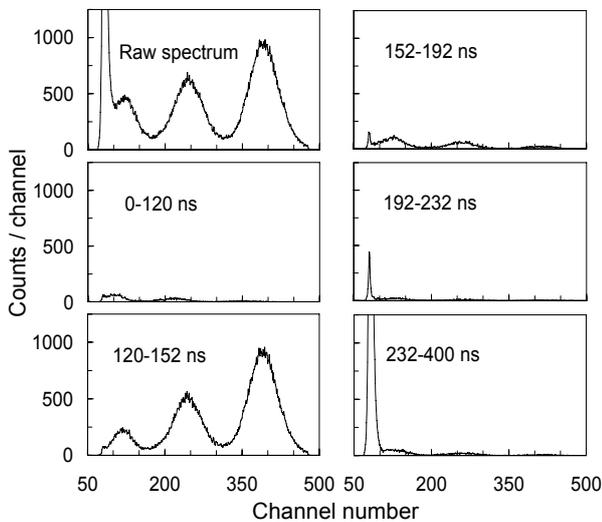

Fig. 40. Raw spectrum and partial pulse-height distributions for different rise-time windows, spanning the full range 0-400 ns, obtained with fluorescence X-rays from silicon and calcium irradiated with a $^{55}$Fe radioactive source.

The main advantage of pulse rise-time discrimination is the significant background reduction. For lower energy X-rays, the effect of rise-time discrimination seems to be more significant because the noise can be efficiently discriminated.

## 6. CONCLUSIONS

We have investigated in detail the performance characteristics of API large area avalanche photodiodes as X-ray and VUV-light detectors. The study has demonstrated that LAAPDs can be used as X-ray detectors in low and medium energy range (up to about 25 keV), being compact and robust detectors with simple operation and very low power consumption. These characteristics make the LAAPDs particularly useful in X-ray spectrometry applications, where pulse rise-time discrimination techniques can be used to reduce the background of X-ray peaks, resulting from pulses which undergo partial amplification in the LAAPD.

The LAAPD performance depends significantly on each individual photodiode since the dark current varies significantly among different LAAPDs. Photodiodes with low dark current are preferred since an energy resolution better than the one of typical proportional counters can be obtained. The LAAPD compact structure, windowless design and high counting rate capability are additional advantages when compared to proportional counters. On the other hand, the limited area and sensitivity to light and temperature are drawbacks.

Good energy linearity of LAAPDs was demonstrated for X-ray energies between 1.7 and 25 keV around the optimum gain. There is however a small effect of non-linearity for higher gains, resulting from space charge effects associated to the local absorption of X-rays. For gains up to 100, the variation of the relative amplitude between signals produced by 22.1 and 5.9 keV X-rays is less than 1%. Another consequence of the point-like absorption of X-rays is the gain fluctuation due to the non-uniform silicon resistivity. The gain non-uniformity effect was measured and 2-3% relative standard deviations of the gain were obtained.

The LAAPD gain and dark current depend strongly on temperature, limiting the LAAPD response in X-ray detection. The relative gain variation with temperature for 5.9 keV X-rays was found to be about -5% per ºC for the highest gains. The LAAPD energy resolution and minimum detectable energy improve with decreasing temperatures down to 0ºC due to the significant reduction in dark current.

LAAPDs were also investigated as VUV photosensors in gas proportional scintillation counters. The relative gain variation with temperature for the VUV light is slightly higher than the one observed for X-rays. As a result of the dark current variation with temperature, both the minimum detection limit and energy resolution improve with decreasing temperature.

The number of electron-hole pairs produced in the LAAPD by light pulses is often determined using X-rays as a reference. This is affected by the gain non-linearity between X-ray and light signals. The LAAPD non-linearity between X-rays and VUV-light pulses was investigated for the first time. The gain ratio between 5.9 keV X-rays and VUV light decreases with gain. For a gain 200, variations of about 10 and 6% were obtained for the argon (~128 nm) and xenon (~172 nm) scintillation, respectively. For comparison, the non-linearity between 5.9 keV X-rays and visible light was also determined. The relative gain variation is now significantly lower (less than 1% for a gain 200). As a result, the non-linearity depends also on the photon penetration depth. VUV photons are superficially absorbed in the LAAPD and, as a consequence, the LAAPD response to VUV-light varies significantly with magnetic field, in opposition to the response to X-rays and visible-light.

ACKNOWLEDGEMENTS

The work was mostly carried out in the Atomic and Nuclear Instrumentation Group of the Instrumentation Centre (Research Unit 217/94) of the Physics Department, University of Coimbra, and received support from Swiss National Science Foundation and Fundação para a Ciência e a Tecnologia (FCT), Portugal, through FEDER and POCI2010 programs, projects POCTI/FIS/13140/98, POCTI/FNU/41720/01, CERN/FIS/43785/01 and POCI/FIS/60534/2004.

L.M.P. Fernandes acknowledges grant from FCT (Ref. SFRH/BD/5426/2001).